\documentclass[final,leqno]{siamltex}

\usepackage{graphicx,amscd,amsmath,amssymb,verbatim,epstopdf,color}

\title{Phase Separation Dynamics in Isotropic Ion-Intercalation  Particles}
\author{Yi Zeng \footnotemark[1]\
\and Martin Z. Bazant\footnotemark[2]}

\begin{document}
\maketitle

\renewcommand{\thefootnote}{\fnsymbol{footnote}}
\footnotetext[1]{Department of Mathematics, Massachusetts Institute of Technology, Cambridge, Massachusetts 02138 \tt(yizeng@mit.edu).}
\footnotetext[2]{Department of Mathematics and Department of Chemical Engineering, Massachusetts Institute of Technology, Cambridge, Massachusetts 02138 \tt(bazant@mit.edu).}

\begin{abstract}
Lithium-ion batteries exhibit complex nonlinear dynamics, resulting from diffusion and phase transformations coupled to ion intercalation reactions. Using the recently developed Cahn-Hilliard reaction (CHR) theory,  we investigate a simple mathematical model of  ion intercalation in a spherical solid nanoparticle, which predicts transitions from solid-solution radial diffusion to two-phase shrinking-core dynamics. This general approach extends previous Li-ion battery models, which either neglect phase separation or postulate a spherical shrinking-core phase boundary, by predicting phase separation only under appropriate circumstances. The effect of the applied current is captured by generalized Butler-Volmer kinetics, formulated in terms of diffusional chemical potentials, and the model consistently links the evolving concentration profile to the battery voltage. We examine sources of charge/discharge asymmetry, such as asymmetric charge transfer and surface ``wetting" by ions within the solid, which can lead to three distinct phase regions.   In order to solve the fourth-order nonlinear CHR initial-boundary-value problem, a control-volume discretization is developed in spherical coordinates. The basic physics are illustrated by simulating many representative cases, including a simple model of the popular cathode material, lithium iron phospate (neglecting crystal anisotropy and coherency strain).  Analytical approximations are also derived for the voltage plateau as a function of the applied current. 
\end{abstract}



\section{Introduction}

The discovery of lithium iron phosphate (Li$_x$FePO$_4$, LFP) as a cathode material for lithium-ion batteries has led to unexpected breakthroughs in the mathematical theory of chemical kinetics coupled to phase transformations~\cite{bazant2013}. Since its discovery in 1997 as a ``low power material" with attractive safety and economic attributes \cite{padhi1997}, LFP has undergone a remarkable reversal of fortune to become the cathode of choice for high-power applications \cite{tarascon2001,kang2009,tang2010}, such as power tools and electric vehicles \cite{ritchie2006, zackrisson2010}, through advances in surface coatings and reduction to nanoparticle form. 

A striking feature of LFP is its strong tendency to separate into stable high density and low density phases, indicated by a wide voltage plateau at room temperature  \cite{padhi1997,tarascon2001} and other direct experimental evidence \cite{delacourt2005,yamada2005,delmas2008,allen2008,oyama2012,chueh2013}.  Similar phase-separation behavior arises in many other intercalation hosts, such as graphite, the typical lithium insertion anode material, which exhibits multiple stable phases.   This has inspired new approaches to model the phase separation process coupled to electrochemistry, in order to gain a better understanding of the fundamental lithium-ion battery dynamics. 

The first mathematical model on two-phase intercalation dynamics in LFP was proposed by Srinivasan and Newman \cite{srinivasan2004}, based on the concept of a spherical ``shrinking core" of one phase being replaced by an outer shell of the other phase, as first suggested by Padhi et al. \cite{padhi1997}.  By assuming isotropic spherical diffusion, the sharp, radial ``core-shell"  phase boundary can be moved in proportion to the current.  This single-particle model was incorporated into traditional porous electrode theory for Li-ion batteries \cite{doyle1993,newman_book} with Butler-Volmer kinetics and concentration dependent diffusivity and fitted to  experiments. The shrinking-core porous-electrode model was recently extended and refitted by Dargaville and Farrell \cite{dargaville2010}. 

In recent years, the shrinking-core hypothesis has been called into question because different phase behavior has been observed experimentally~\cite{laffont2006,chen2006,allen2008,delmas2008,chueh2013} and predicted theoretically~\cite{bazant2013}. It has become clear that a more realistic particle model must account for two-phase thermodynamics~\cite{han2004,singh2008,lai2010,lai2011b,zeng2013MRS}, crystal anisotropy~\cite{singh2008,bai2011,tang2011}, coherency strain~\cite{cogswell2012}, surface energy~\cite{cogswell2013}, and reaction limitation in nanoparticles~\cite{singh2008,bai2011,bai2014}, and electrochemical interactions between large numbers of such particles in porous electrodes~\cite{ferguson2012,bai2013,ferguson2014,orvananos2014}. In  larger, micron-sized particles, the shrinking-core model may still have some relevance  due to solid diffusion limitation and defects (such as dislocations and micro cracks) that can reduce coherency strain~\cite{singh2008,burch2009,dargaville2013CHR}.  Moreover, diffusion becomes more isotropic in larger particles due to the increased frequency of point defects, such as channel-blocking Fe anti-site defects  in LFP~\cite{malik2010}. 

Regardless of the details of the model, fundamental questions remain about the dynamics of phase separation driven by electrochemical reactions, even in the simplest case of an isotropic strain-free spherical particle. When should we expect core-shell phase separation versus pure diffusion in a solid solution? What other transient phase morphologies are possible? How are reaction kinetics affected by phase separation?  Traditional battery models, which place artificial spherical phase boundaries and assume classical Butler-Volmer kinetics, are not able to answer these questions.

In this article, we formulate a simple mathematical model that captures the essential features of {\it bulk} phase separation coupled to Faradaic intercalation reactions in a single solid nanoparticle. The model is  based on a recently developed mathematical theory of chemical reaction and charge transfer kinetics based on nonequilibrium thermodynamics~\cite{bazant2013}, which we review in Section ~\ref{sec:back}.  In the case of an isotropic, strain-free spherical particle, the resulting Cahn-Hilliard reaction (CHR) equations are formulated for Butler-Volmer (BV) kinetics and regular solution thermodynamics in Section ~\ref{sec:eqns}. The model predicts smooth concentration profiles limited by radial diffusion with smooth voltage profiles versus state of charge in cases of solid-solution thermodynamics (Section ~\ref{sec:ss}) and radial phase separation with a flat  voltage plateau in cases of two stable phases (Section ~\ref{sec:phasesep}), which are strongly affected by surface wetting (Section ~\ref{sec:wet}).  After summarizing the results, in Section~\ref{sec:num} we present the control-volume numerical scheme for the CHR model that allows us to accurately solve this stiff fourth-order nonlinear initial-boundary-value problem.

\section{Background}
\label{sec:back}

A systematic approach to describe chemical kinetics coupled to phase transformations has recently been developed by Bazant ~\cite{bazant2013}, based on nonequilibrium thermodynamics.  The theory leads to a general reaction-diffusion equation of the form,
\begin{equation}
\frac{\partial c_i}{\partial t} = \nabla \cdot \left( M_i c_i \nabla \frac{\delta G}{\delta c_i} \right) + R_i\left( \left\{ \frac{\delta G}{\delta c_j}\right\} \right)  \label{eq:rd}
\end{equation}
where $c_i$ is the concentration, $M_i$ the mobility,  and $R_i$ the volumetric reaction rate of species $i$, assuming homogeneous kinetics. The diffusive flux (second term) and the reaction rate (third term) are both expressed in terms of diffusional chemical potentials,
\begin{equation}
\mu_i = \frac{\delta G}{\delta c_i}   \label{eq:mudef}
\end{equation}
defined as variational derivatives of the total free energy functional $G[\{c_i\}]$.  Physically, $\mu_i(x)$ is free energy required to add a continuum ``particle" (delta function) of species $i$ to the system at position $x$.  

For the conversion of reactants $\{ A_r\}$ to products $\{ B_p \}$, \textcolor{black}{given that the stoichiometric coefficients are $ s_r$ and $ s_p$ for reactants and products, respectively},
\begin{equation}
 \sum_r s_r \mbox{A}_r \to  \sum_p s_p \mbox{B}_p,  \label{eq:genreact}
\end{equation}
assuming thermally activated kinetics,  the  reaction rate has the general variational form,
\begin{equation}
R = \frac{k_0}{\gamma_\ddag}   \left[ \exp\left( \sum_r \frac{s_r}{k_BT} \frac{\delta G}{\delta c_r}\right) - \exp\left( \sum_p \frac{s_p}{k_BT} \frac{\delta G}{\delta c_p} \right)\right]    \label{eq:Rphase}
\end{equation}
where $\gamma_\ddag$ is the activity coefficient of the transition state and $R_i = \pm s_i R$ ($+$ for products, $-$ for reactants).  A mathematical model of the general form (\ref{eq:rd}) was perhaps first proposed by Hildebrand {\it et al.} to describe nanoscale pattern formation in catalytic surface reactions~\cite{hildebrand1999,hildebrand2003} and corresponds to specific models for the free energy ($G$) and the transition state ($\gamma_\ddag$).  In the case of electrochemical reactions involving ions and electrons, different assumptions that also account for electrostatic energy lead to Bazant's generalizations of the classical Butler-Volmer and Marcus theories of charge transfer for concentrated solutions and solids~\cite{bazant2013}. \textcolor{black}{  Fehribach and O'Hayre~\cite{fehribach2009} and Lai and Cuicci~\cite{lai2010,lai2011a} have also recently recast the Butler-Volmer equation in terms of electrochemical potentials, but without relating the exchange current to chemical activities or using the general variational formulation (\ref{eq:mudef}).  }

The variational reaction-diffusion equation (\ref{eq:rd}) unifies the Cahn-Hilliard and Allen-Cahn equations from phase-field modeling in a general formulation of non-equilibrium chemical thermodynamics for reacting mixtures. These classical equations, widely used in materials science and applied mathematics~\cite{kom}, are special cases of Eq. (\ref{eq:rd}) that correspond to rate limitation by diffusion,
\begin{equation}
\frac{\partial c}{\partial t} = \nabla \cdot \left( M c \nabla \frac{\delta G}{\delta c} \right)  \ \ \ \mbox{(Cahn-Hilliard)}
\end{equation}
or by linear reaction kinetics for a small thermodynamic driving force, 
\begin{equation}
\frac{\partial c}{\partial t} = - k  \frac{\delta G}{\delta c}  \ \ \ \mbox{ (Allen-Cahn)}
\end{equation}
respectively~\cite{singh2008,bazant2013}.  The general equation (\ref{eq:rd}) can be applied to many problems in chemical or electrochemical dynamics~\cite{bazant2013}. In the case of ion intercalation in Li-ion battery nanoparticles, it has mainly been studied in two limiting cases.  

For reaction-limited anisotropic nanoparticles, the general theory can be reduced to the Allen-Cahn reaction (ACR) equation,
\begin{equation}
\frac{\partial c}{\partial t} = R\left( \left\{ \frac{\delta G}{\delta c}\right\} \right) \ \ \ \mbox{(ACR)}
\end{equation}
for the depth-averaged ion concentration $c(x,y)$ along the active surface where intercalation reactions occur, 
as shown by Bai et al.~\cite{bai2011} and Burch~\cite{burch_thesis}, building on the seminal paper of Singh et al.~\cite{singh2008}. The ACR model  has been applied successfully to predict experimental data for LFP, using generalized Butler-Volmer kinetics and accounting for coherency strain, by Cogswell and Bazant~\cite{cogswell2012,cogswell2013,bazant2013}. An important prediction of the ACR model is the dynamical suppression of phase separation at high rates~\cite{bai2011,cogswell2012}, as it becomes favorable to spread reactions uniformly over the particle surface, rather than to focus them on a thin interface between stable phases.  The ACR model has also been used to predict a similar transition in electrochemical deposition of Li$_2$O$_2$ in Li-air battery cathodes, from  discrete particle growth at low currents to uniform films at high currents~\cite{horstmann2013}.

For larger particles, the Cahn-Hilliard reaction (CHR) model, 
\begin{equation}
\frac{\partial c}{\partial t} + \nabla\cdot\mathbf{F} = 0, \ \  \mathbf{F} = -  M c \nabla \frac{\delta G}{\delta c_i}, \ \ 
-\hat{n}\cdot \mathbf{F} = R\left( \left\{ \frac{\delta G}{\delta c}\right\} \right)  \ \ \ \mbox{(CHR)}
\label{eq:chr}
\end{equation}
describes bulk phase separation driven by heterogenous reactions, which are localized on the surface and described by a flux matching boundary condition~\cite{bazant2013}.  This general model was first posed by Singh, Ceder and Bazant \cite{singh2008} but received less attention until recently.  For Butler-Volmer kinetics, Burch and Bazant~\cite{burch2009,burch_thesis} and Wagemaker et al.~\cite{wagemaker2011} solved the CHR model in one dimension to describe size-dependent miscibility in nanoparticles.  Dargaville and Farrell~\cite{dargaville2013CHR,dargaville_thesis} first solved \textcolor{black}{the} CHR in two dimensions (surface and bulk) for a rectangular particle using a least-squares based finite-volume method~\cite{dargaville2013numerical} and examined the transition to ACR behavior with increasing crystal anisotropy and surface reaction limitation. They showed that phase separation tends to persist within large particles, similar to the shrinking core picture, if it is not suppressed by coherency strain and/or fast diffusion perpendicular to the most active surface.

\section{Cahn-Hilliard Reaction Model}
\label{sec:eqns}

In this work, we solve the CHR model with generalized Butler-Volmer kinetics for a spherical host particle with the intercalated ion concentration varying only in the radial direction.  Spherical symmetry is also the most common approximation for solid diffusion in traditional Li-ion battery models~\cite{doyle1993,zeng2013numerical}.  This simple one-dimensional version of the CHR model is  valid for large, defective crystals with negligible coherency strain and isotropic diffusion~\cite{singh2008,burch_thesis,dargaville2013CHR,dargaville_thesis}. It may also be directly applicable to low-strain materials such as lithium titanate~\cite{ozhuku1995}, a promising long-life anode material~\cite{yang2009}.   We simulate phase separation dynamics at constant current, which sometimes, but not always, leads to shrinking-core behavior.  Related phase-field models of isotropic spherical particles, including the possibility of simultaneous crystal-amorphous transitions, have also been developed and applied to LFP by Tang et al.~\cite{tang2009,tang2010}, Meethong et al.~\cite{meethong2007,meethong2007a,meethong2008}, and Kao et al~\cite{kao2010}, but without making connections to charge-transfer theories from electrochemistry.  Here, we focus on the electrochemical signatures of different modes of intercalation dynamics -- voltage transients at constant current -- which are uniquely provided by the CHR model with consistent Butler-Volmer reaction kinetics~\cite{bazant2013}. We also consider the nucleation of phase separation by surface wetting~\cite{bai2011}, in the absence of coherency strain, which would lead to a size-dependent nucleation barrier~\cite{cogswell2013} and symmetry-breaking striped phase patterns~\cite{vanderven2009,cogswell2012}.

\subsection{ Model formulation}
Consider the CHR model (\ref{eq:chr}) for a spherical, isotropic, strain-free, electron-conducting particle of radius $R_p$ with a concentration profile $c(r,t)$ of intercalated ions (number/volume). As first suggested by Han et al. for LFP~\cite{han2004}, we assume the chemical potential of the Cahn-Hilliard regular solution model~\cite{cahn1958,cahn1959-1,cahn1959-2},
\begin{equation}
\mu = k_B T \ln \left(\frac{c}{c_m-c}\right) + \Omega \left(\frac{c_m-2c}{c_m}\right) - \frac{\kappa }{c_m^2} \nabla^2c,
\end{equation}
where $k_B$ is Boltzmann's constant, $T$  the absolute temperature, $\Omega$  the enthalpy of mixing per site, $\kappa$  the gradient energy penalty coefficient, $V_s$ the volume of each intercalation site, and $c_m=V_s^{-1}$ is the maximum ion density. Although we account for charge transfer at the surface (below), we set the bulk electrostatic energy to zero, based on the assumption each intercalated ion diffuses as a neutral polaron, coupled to an adjacent mobile electron, e.g. reducing a metal ion such as Fe$^{3+}+e^-\to$ Fe$^{2+}$ in LFP.  (For semiconducting electrodes, imbalances in ion and electron densities lead to diffuse charge governed by Poisson's equation in the CHR model~\cite{bazant2013}.)

The mobility $M$ in the flux expression (\ref{eq:chr}) is related to the tracer diffusivity $D$ by the Einstein relation, $D = M k_BT$.
For thermodynamic consistency with the regular solution model, the tracer diffusivity must take into account excluded sites
\begin{equation}
D = D_0 \left( 1 - \frac{c}{c_m}\right) = M k_BT
\end{equation}
where $D_0$ is the dilute-solution limit, which leads to the ``modified Cahn-Hilliard equation"~\cite{nauman2001}.  This form also follows consistently from our reaction theory, assuming that the transition state for solid diffusion excludes two sites~\cite{bazant2013}.

At the surface of the particle, $R=R_p$, the insertion current density $I(t)$ is related to the voltage $V(t)$ and surface flux density $ F(R_p,t)$, where $\mathbf{F} = F \hat{R}$ is the radial flux. By charge conservation, the current is the integral of the surface flux times the charge per ion $ne$,
\begin{equation}
\label{eqn:ChargeConservationCondition}
I = - n e F(R_p,t),
\end{equation}
where $e$ is the electron charge. Electrochemistry enters the model through the current-voltage relation, $I(V,c,\mu)$, which depends on $c$ and $\mu$ at the surface.  Here, we adopt thermodynamically consistent, generalized Butler-Volmer kinetics for the charge-transfer rate~\cite{bazant2013}, given below in dimensionless form.  

We also impose the ``natural" or ``variational" boundary condition for the fourth-order Cahn-Hilliard equation,
\begin{equation}
\frac{\partial c}{\partial r}(R_p,t) = c_m^2 \frac{\partial \gamma_s}{\partial c},
\end{equation}
where $\gamma_s(c)$ is the surface energy per area, which generally depends on ion concentration.  The natural boundary condition expresses continuity of the chemical potential and controls the tendency for a high or low concentration solid phase to preferentially ``wet" the surface from the inside~\cite{cahn1977,cogswell2013}. Together with symmetry conditions, $F(0,t)=0$ and $\frac{\partial c}{\partial R}(0,t)=0$, we have the required four boundary conditions, plus the current-voltage relation, to close the problem.

\subsection{ Dimensionless equations}
To nondimensionalize the system, we will use several basic references to scale the model, which include the particle radius $R_p$ for the length scale, the diffusion time $\frac{R_p^2}{D_0}$ for the time scale, the maximum ion concentration $c_m$ for the concentration scale and the thermal energy $k_BT$ for any energy scale. The dimensionless variables are summarized in Table \ref{table:1}.

\begin{table}[!h]
\caption{Dimensionless variables in the CHR model.}
\label{table:1}
\centering
\begin{tabular}{  c c c c c } \hline
$\tilde{c} = \frac{c}{c_m}$ & $ \tilde{t} = \frac{  D_0 }{R^2_p} t$ & $\tilde{r} = \frac{r}{R_p}$  &
$\tilde{\nabla} = R_p \nabla$ & $\tilde{F} = \frac{R_p}{c_m D_0} F$ \\ 
  $\tilde{\mu} = \frac{\mu}{k_B T}$ &   $\tilde{\Omega} = \frac{\Omega}{k_BT}$  
  & $\tilde{\kappa} = \frac{\kappa }{R_p^2 c_m k_BT}$  &
 $\tilde{I} = \frac{  R_p }{c_m  ne  D_0} I$ & $\tilde I_0 = \frac{  R_p }{c_m  ne D_0} I_0$  \\
 $\tilde{\eta} = \frac{e}{k_B T} \eta$  &
  $\tilde{V} = \frac{eV}{k_BT}$ &  $\tilde{V}^\Theta = \frac{eV^\Theta}{k_BT}$ &
    $\tilde \gamma_s = \frac{ \gamma_s}{R_p c_m k_B T} $  & $\beta =  \frac{1}{\tilde{\kappa} } \frac{\partial \tilde  \gamma_s}{\partial \tilde c}$ \\
  \hline
\end{tabular}
\end{table}

With these definitions, our model takes the dimensionless form,
\begin{eqnarray}
\label{eqn:MassConservation}
\frac{\partial \tilde c}{\partial \tilde t} = - \frac{1}{\tilde{r}^2} \frac{\partial}{\partial \tilde{r}}
\left( \tilde{r}^2  \tilde{F} \right) \\
\tilde F = - (1-\tilde c) \tilde c \frac{ \partial \tilde \mu}{\partial \tilde{r}} \\
\tilde \mu = \ln \frac{\tilde c}{1- \tilde c} + \tilde \Omega (1-2\tilde c) - \tilde \kappa \tilde \nabla^2 \tilde c \\
\frac{\partial \tilde{c}}{\partial \tilde r}(0,\tilde{t}) = 0, \ \ \ 
\frac{\partial \tilde{c}}{\partial \tilde r}(1,\tilde{t}) = \beta \\
\tilde  F(0,\tilde{t}) = 0, \ \ \ 
\tilde  F(1,\tilde{t})  = \textcolor{black}{\tilde I}.
\end{eqnarray}
In order to relate the current to the battery voltage, we assume generalized Butler-Volmer kinetics~\cite{bazant2013}, 
\begin{eqnarray}
\tilde I &=& \tilde I_0 \left( e^{-\alpha \tilde  \eta}-e^{(1-\alpha) \tilde  \eta} \right) \\
\tilde{\eta} &=& \tilde  \mu+  \tilde V - \tilde{V}^\Theta \\
\tilde I_0 &=& \tilde c^{\alpha}(1-\tilde c)^{1-\alpha}  e^{\alpha (\tilde \Omega (1-2\tilde c) - \tilde \kappa \nabla^2 \tilde c) }= (1-\tilde c) e^{ \alpha \tilde \mu}
\end{eqnarray}
where \textcolor{black}{$\tilde I$} is the \textcolor{black}{nondimensional} insertion current density (per area),   \textcolor{black}{$\tilde I_0$} the \textcolor{black}{nondimensional} exchange current density,  $\alpha$ the charge transfer coefficient, \textcolor{black}{$\tilde \eta$} the \textcolor{black}{nondimensional} surface or activation overpotential,  \textcolor{black}{$\tilde V$} the \textcolor{black}{nondimensional} battery voltage, and \textcolor{black}{$\tilde V^\Theta$} the \textcolor{black}{nondimensional} reference voltage for a given anode (e.g. Li metal) when the particle is homogeneous at $\tilde{c}=\frac{1}{2}$.  The derivation of this rate formula assumes that the transition state for charge transfer excludes one surface site, has no enthalpic excess energy, and has an electrostatic energy $(1-\alpha)$ times that of the electron plus the ion in the electrolyte.  It is common to assume $\alpha=\frac{1}{2}$, but we will relax this assumption below.   In equilibrium, \textcolor{black}{$\tilde \eta=0$}, the interfacial voltage, \textcolor{black}{$\Delta \tilde V = \tilde V - \tilde V^\Theta$} is determined by the Nernst equation, $\Delta\tilde{V}_{eq} = -\tilde{\mu}$.    Out of equilibrium, the overpotential, \textcolor{black}{$\tilde \eta(t) = \Delta \tilde V(t) - \Delta \tilde V_{eq}(t)$}, is determined by solving for the transient concentration profile.

\subsection{Governing parameters} 
Dimensionless groups are widely used in fluid mechanics to characterize dynamical regimes~\cite{barenblatt_book}, and recently the same principles have been applied to intercalation dynamics in Li-ion batteries~\cite{singh2008,ferguson2012}.
The CHR model is governed by four dimensionless groups, $\tilde{\Omega}$, $\tilde{\kappa}$,  $\beta$ and $\tilde{I}$ (or $\tilde{V}$) with the following physical interpretations. 

The ratio of the regular solution parameter (enthalpy of mixing) to the thermal energy can be positive or negative, but in the former case (attractive forces) it can be interpreted as
\begin{equation}
\tilde \Omega=  \frac{\Omega}{k_BT} = \frac{2 T_c}{T},
 \end{equation}
i.e. twice the ratio of the critical temperature $T_c=\frac{\Omega}{2k_B}$, below which phase separation is favored,  to the temperature $T$.   Below the critical point, $T<T_c$ (or $\tilde \Omega > 2$), the thickness and interfacial tension of the diffuse phase boundary scale as $\lambda_b=\sqrt{\kappa / c_m \Omega}$ and $\gamma_b=\sqrt{\kappa \Omega c_m}$, respectively~\cite{cahn1958}, so the dimensionless gradient penalty 
\begin{equation}
\tilde{\kappa} = \frac{\kappa }{c_m k_BT R_p^2} = \tilde \Omega \left( \frac{\lambda_b}{R_p} \right)^2 \ll 1
\end{equation}
equals $\tilde\Omega$ times the squared ratio of the interfacial width (between high- and low-density stable phases) to the particle radius, which is typically small.

The parameter $\beta$  is the dimensionless concentration gradient at the particle surface, 
$\beta =  \frac{1}{\tilde{\kappa} } \frac{\partial \tilde  \gamma_s}{\partial \tilde c}$, which we set to a constant, assuming that  the surface tension $\gamma_s(c)$ is a linear function of composition.  Letting $\Delta \gamma_s = \frac{\partial \gamma_s}{\partial \tilde{c}}$ be the difference in surface tension between high-density $(\tilde{c}\approx 1)$ and low-density  $(\tilde{c}\approx 1)$ phases, 
\begin{equation}
\beta= \frac{R_p}{\lambda_b} \frac{\Delta \gamma_s}{\gamma_b}  \gg 1
\end{equation}
we can interpret $\beta$ as the  ratio of particle size to the phase boundary thickness times the surface-to-bulk phase boundary tension ratio, $\frac{\Delta \gamma_s}{\gamma_b}$. In cases of partial ``wetting" of the surface by the two solid phases, this  ratio is related to the equilibrium contact angle $\theta$ by Young's Law, 
\begin{equation}
\cos\theta = \frac{\Delta \gamma_s}{\gamma_b}.
\end{equation}
Partial wetting may occur in the absence of elastic strain (as we assume below), but complete wetting by the lower-surface-energy phase is typically favored for coherent phase separation because $\gamma_b \ll |\Delta \gamma_s|$~\cite{cogswell2013}.  In any case, for thin phase boundaries, we typically have $\beta \gg 1$.

Finally, the current density is scaled to the diffusion current,
\begin{equation}
\tilde{I}  =  \frac{I  }{3ne c_m V / (\tau_D A)}= \frac{R_p}{ne c_m D_0} I,
\end{equation}
where $V = \frac{4}{3} \pi R^3$ is the volume of the sphere, $ne c_m V$ represents the maximum charge \textcolor{black}{that} can be stored in the sphere, $A= 4\pi R_p^2$ is the surface area and $\tau_D = R_p^2/D_0$ is the diffusion time into the particle. $\tilde{I} = 1$ is equivalent to \textcolor{black}{the particle that} can be fully charged from empty in $\frac{1}{3}$ unit of diffusion time $\tau_D$ with this current density. The exchange current has the same scaling. Rate limitation by surface reactions or by bulk diffusion corresponds to the limits $\tilde{I}_0 \ll 1$ or $\tilde{I}_0 \gg 1$, respectively, so this parameter behaves like a Damkoller number~\cite{singh2008,ferguson2012}.

\subsection{ Simulation details}

For a given dynamical situation, either the current or the voltage is controlled, and the other quantity is predicted by the model.  Here we consider the typical situation of ``galvanostatic" discharge/charge cycles at constant current, so the model predicts the voltage $V$, which has the dimensionless form, $\tilde{V} = \frac{neV}{k_BT}$.   The electrochemical response is typically plotted as voltage versus state of charge, or mean filling fraction,
\begin{equation}
X =\frac{ \int c \, dV }{\frac{4}{3} \pi R_p^3 c_{m}}.
\end{equation}
The reference scale for all potentials is the thermal voltage, $\frac{k_BT}{e}$, equal to 26 mV at room temperature.

\begin{table}[!h]
\caption{Parameter settings for LFP~\cite{cogswell2012,cogswell2013} used in the numerical simulations, except as otherwise noted.}
\label{table:2}
\centering
\begin{tabular}{  c c c | c c c}
	\hline
Parameter & Value & Unit  & Parameter & Value & Unit \\ \hline
$R_p$                                                         & $1 \times 10^{-7}$                            & m  &
$\Omega$                                                  & $0.115$                           & eV \\
$\kappa$                                                    & $3.13 \times 10^{9}$                         & eV/m  &
$D_0$                                                 & \textcolor{black}{$1 \times 10^{-14}$}                       & m$^2$/s \\
$c(r,0)$                                               & $10$                                                    & mol / m$^3$ &
$c_{m}$                                                & $1.379 \times 10^{28}$                       & m$^-3$ \\
$n$                                                             & $1$                                                        & - &   
$\alpha$                                                     & $0.5$                                                   & - \\
$V^\Theta $                                         &  $3.42$                                                 & V &
$I_0$                                                  & \textcolor{black}{$1.6\times 10^{-4}$}                        & A/m$^2$\\
     \hline
\end{tabular}
\end{table}

In the following sections, we perform numerical simulations for the parameter settings in Table \ref{table:2}, which  have been fitted to experimental and {\it ab initio} computational results for LFP \cite{malik2010, bai2011, cogswell2012, bai2014},  but we vary $\tilde \Omega$ to obtain different dynamical behaviors, which may represent other Li-ion battery materials. \textcolor{black}{ Reports of the lithium diffusivity in the solid vary widely in the literature and reflect different modeling approaches. Fits of the shrinking core model to experimental data yield $D=8\times 10^{-18}$ ~\cite{srinivasan2004,dargaville2010}, but this is five orders of magnitude smaller than the anisotropic perfect-crystal diffusivity $D_b\approx 10^{-12}$ m$^2$/s along the fast $b$ axis predicted by {\it  ab initio} calculations~\cite{morgan2004}. Here, we use the value $D=10^{-14}$ m$^2$/s, as predicted for a 1\% density of Fe anti-site defects blocking the $b$-axis channels in particles of size 0.1-1.0 $\mu$m \cite{malik2010}, which also lead to more isotropic diffusion.   }

\textcolor{black}{ Even larger discrepancies for the exchange current density arise in the battery literature. This is partly due to different surface coatings on the active particles, but there is clearly also   a need for improved mathematical models to fit experimental data, since charge-transfer reaction rates are difficult to calculate from first principles.  Fits to the shrinking core model yield $I_0 = 3\times 10^{-6}$ \cite{srinivasan2004} or $5\times 10^{-5}$ A/m$^2$ \cite{dargaville2010}, but much larger values up to  $19$ A/m$^2$ have also been reported ~\cite{cwang2007,kasavajjula2008}.  Here, we use the intermediate value, ${I}_0 = 1.6\times 10^{-4}$ A/m$^2$ ~\cite{bai2014}, obtained from experiments by fitting to a simple model of composite-electrode phase transformation dynamics~\cite{bai2013}, assuming a uniform reaction rate over each particle, as we do in our isotropic model here. (Larger local values of $I_0$ per surface site would be implied by inhomogeneous filling, e.g. by intercalation waves~\cite{singh2008,bai2011,cogswell2012,cogswell2013}.  The same study also quantitatively supports the Marcus-Hush-Chidsey theory of charge transfer from the carbon coating to the solid~\cite{bai2014}, but here we adopt the simpler Butler-Volmer equation used in all of battery models.) } 

\textcolor{black}{ 
In our simulations, we consider a typical active nanoparticle of size $R_p = 100$ nm. Using the parameters above for LFP, solid diffusion is relatively fast,} allowing us to focus on the novel coupling of reaction kinetics with phase separation~\cite{bazant2013}.   \textcolor{black}  In this exercise, we initially neglect surface wetting (by setting $\beta=0$) and coherency strain, both of which are important for an accurate description of LFP~\cite{cogswell2012,cogswell2013}. In later sections, we also consider $\beta > 0$ and $\alpha \neq \frac{1}{2}$ for the more interesting cases of phase separation ($\tilde\Omega > 2$).  We employ a control volume method (described below) for the spatial discretization of the system and the ode15s solver in MATLAB for the time integration.  { Consistent with common usage, we report the total current in terms of the ``C-rate", C/$n$, which means full charge or discharge (i.e. emptying or filling) of the particle in $n$ hours; for example, ``C/10" and ``10C" mean full discharge in 10 hours or 6 minutes, respectively.    }

\section{Solid Solution}
\label{sec:ss}

Our model predicts simple diffusive dynamics with slowly varying concentration and voltage transients under ``solid solution" conditions, where configurational entropy promotes strong mixing.  The regular solution model predicts that bulk solid solution behavior occurs at all temperature if there are repulsive forces between intercalated ions, $\Omega < 0$, or above the critical temperature $T > T_c$ for attractive ion-ion forces, $\Omega > 0$. Here, we  consider finite-sized particles and examine current-voltage transients in both of these cases of solid-solution thermodynamics. 

\subsection{Repulsive forces}

\begin{figure}[!h]
\centering
\begin{tabular}{ c c}
\includegraphics[scale=0.43]{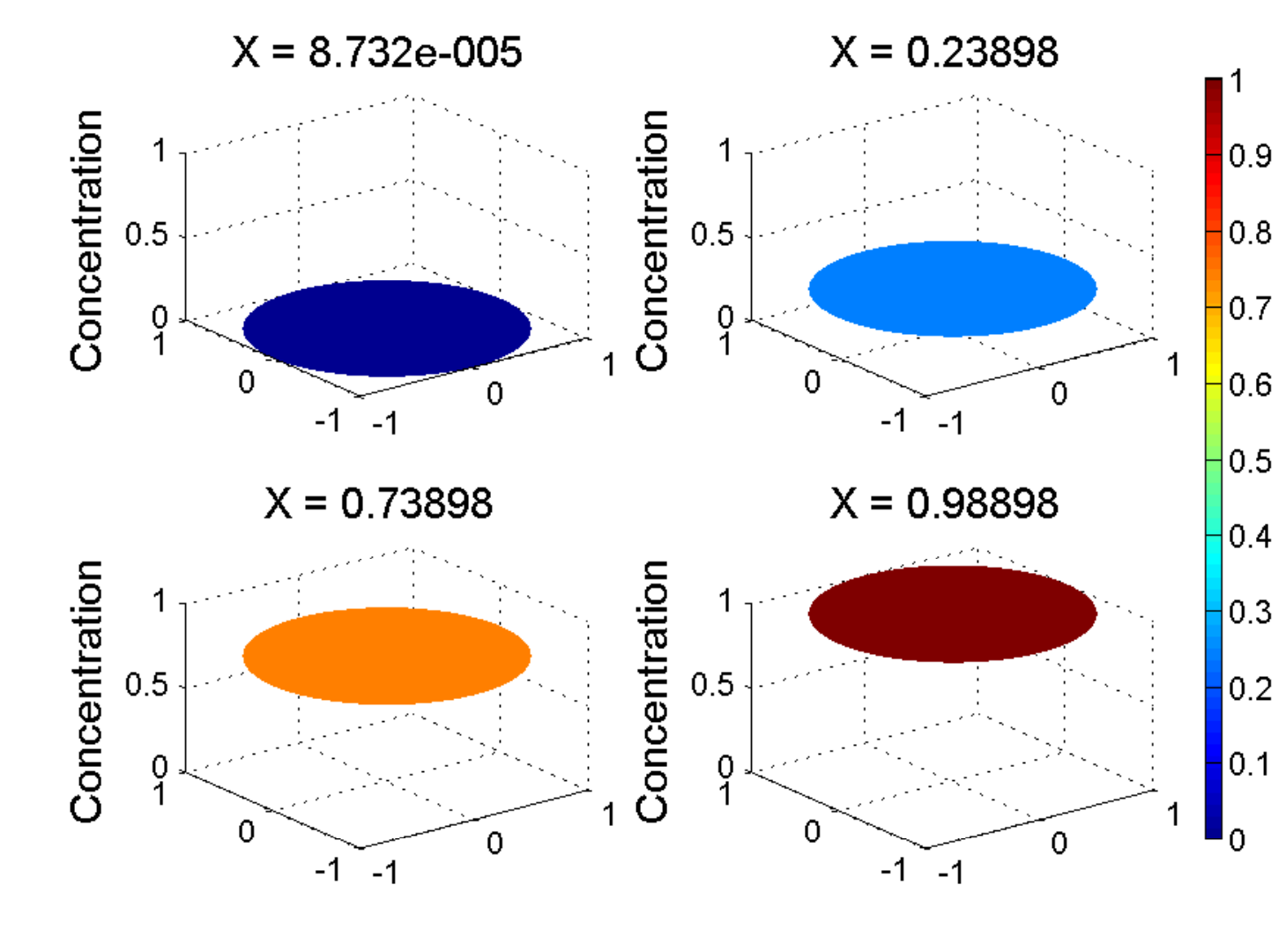} & \includegraphics[scale=0.43]{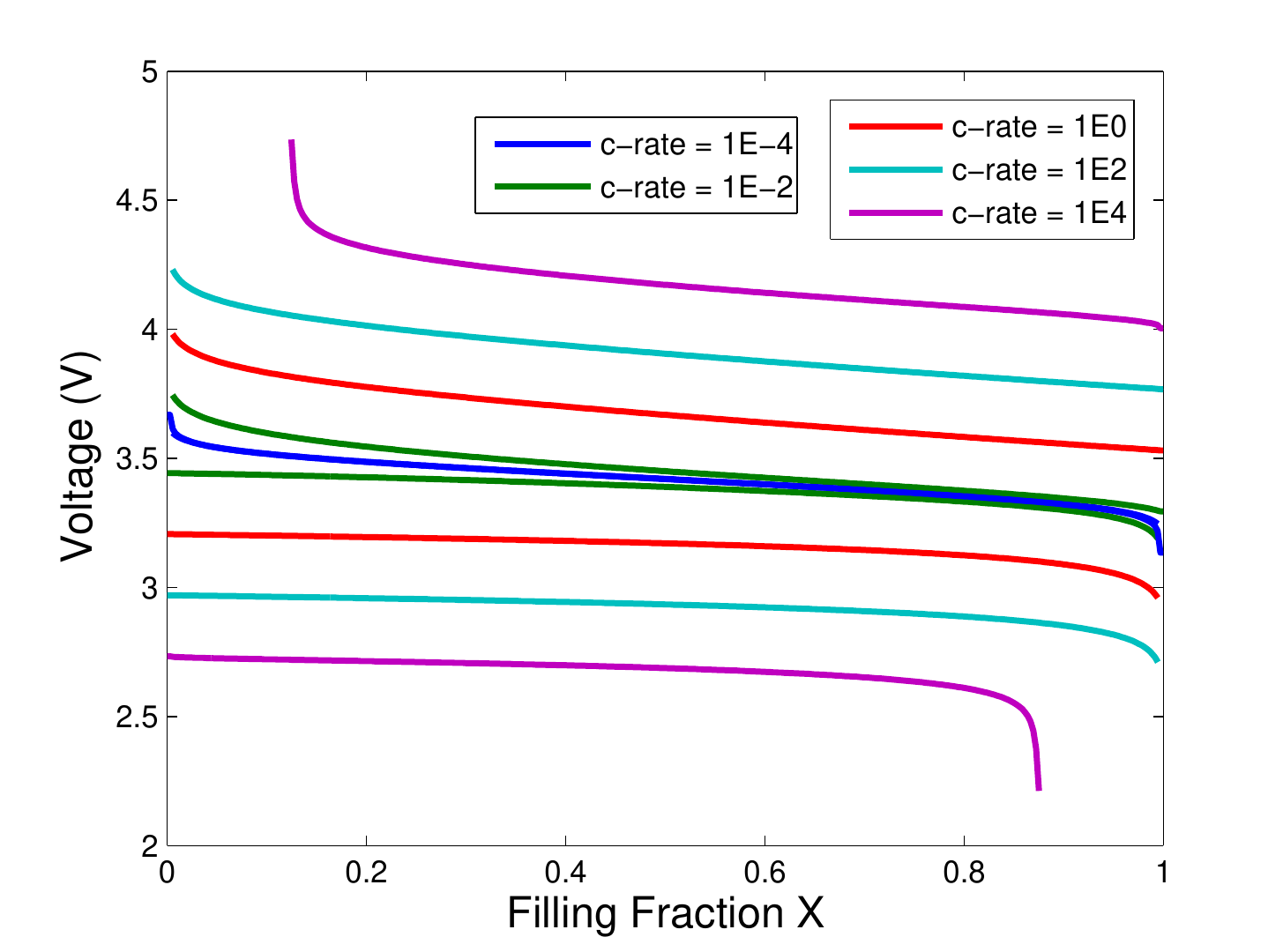}\\   
\end{tabular}
\caption{ Constant current cycling of a spherical intercalation particle, composed of a solid solution of lithium ions with repulsive forces ($\tilde \Omega =  -2$).  Left: profiles of dimensionless concentration $\tilde{c}(\tilde{r})$ (local filling fraction) at different mean compositions (average filling fraction, $X$) at \textcolor{black}{constant current $C/1$}.  The vertical dimension in the plots shows the concentrations, while the horizontal circle denotes the \textcolor{black}{planar cross section} at the equator of the sphere. Right:  voltage versus state of charge (filling fraction) at different currents.  \textcolor{black}{The ten voltage curves represent C-rates of $ = \pm 10^{-4}$C$, \pm 10^{-2}$C$, \pm 10^{0}$C$, \pm 10^{2}$C$, \pm 10^{4}$C.} }
\label{fig:NegativeOmega}
\end{figure}

A negative enthalpy of mixing, $\Omega < 0$, reflects mean-field attraction between ions and vacancies, or equivalently, repulsion between intercalated ions that promotes homogeneous intercalation.  Consider galvanostatic (constant current) charge and discharge cycles with $\Omega = -0.0514$eV or  $\tilde \Omega=-2$. When the current is small, $\tilde{I}\ll 1$, diffusion is fast, and the ions remain uniformly distributed inside the particle during intercalation dynamics, as shown in Fig. \ref{fig:NegativeOmega}.  At high currents, $\tilde{I}\gg 1$ (not considered here), diffusion becomes rate limiting, and concentration gradients form, as in prior models of spherical nonlinear diffusion~\cite{doyle1993,srinivasan2004,zeng2013numerical}.

Given the Butler-Volmer symmetry factor, $\alpha = 0.5$, and assuming uniform composition, the total voltage drop between anode and particle surface is given by 
\begin{equation}
\tilde{V} = \tilde{V}^{\Theta} -\tilde \mu(\tilde c) - 2 \sinh^{-1} \left(\frac{\tilde I}{2 \tilde I_0(\tilde c)}\right), \label{eq:V1}
\end{equation}
where $V$ is the battery voltage, $V^{\Theta}$ is the constant reference voltage for a given anode, and $\tilde I_0(\tilde c)$ the exchange current density at the given concentration profile.  The simulated discharge curves in Fig. \ref{fig:NegativeOmega} fit this expression well and exhibit no voltage plateau (a signature of phase separation discussed below).  The model exhibits a positive internal resistance, since the battery voltage decreases for $I>0$ (discharging) and increases for $I<0$ (charging).  According to Eq. (\ref{eq:V1}), the voltage increment, or overpotential, has two sources:  concentration changes at the surface that shift the Nernst equilibrium interfacial voltage (second term, concentration overpotential) and Butler-Volmer charge-transfer resistance (third term, activation overpotential).

\subsection{Weak attractive forces or high temperature}

When the mixing enthalpy per site $\Omega$ is positive, there is an effective repulsion between ions and vacancies, or equivalently, an attraction between ions that promotes phase separation into Li-rich and Li-poor phases.  This tendency \textcolor{black}{is} counteracted by configurational entropy, which always promotes the mixing of ions and vacancies and leads to homogeneous solid solution behavior at high temperature $T$.  Below the critical temperature, $T< T_c=\frac{\Omega}{2 k_B}$, attractive forces overcome configurational entropy, leading to stable bulk  phase separation.  

\begin{figure} [!h]
\begin{center}
\includegraphics[scale=0.5]{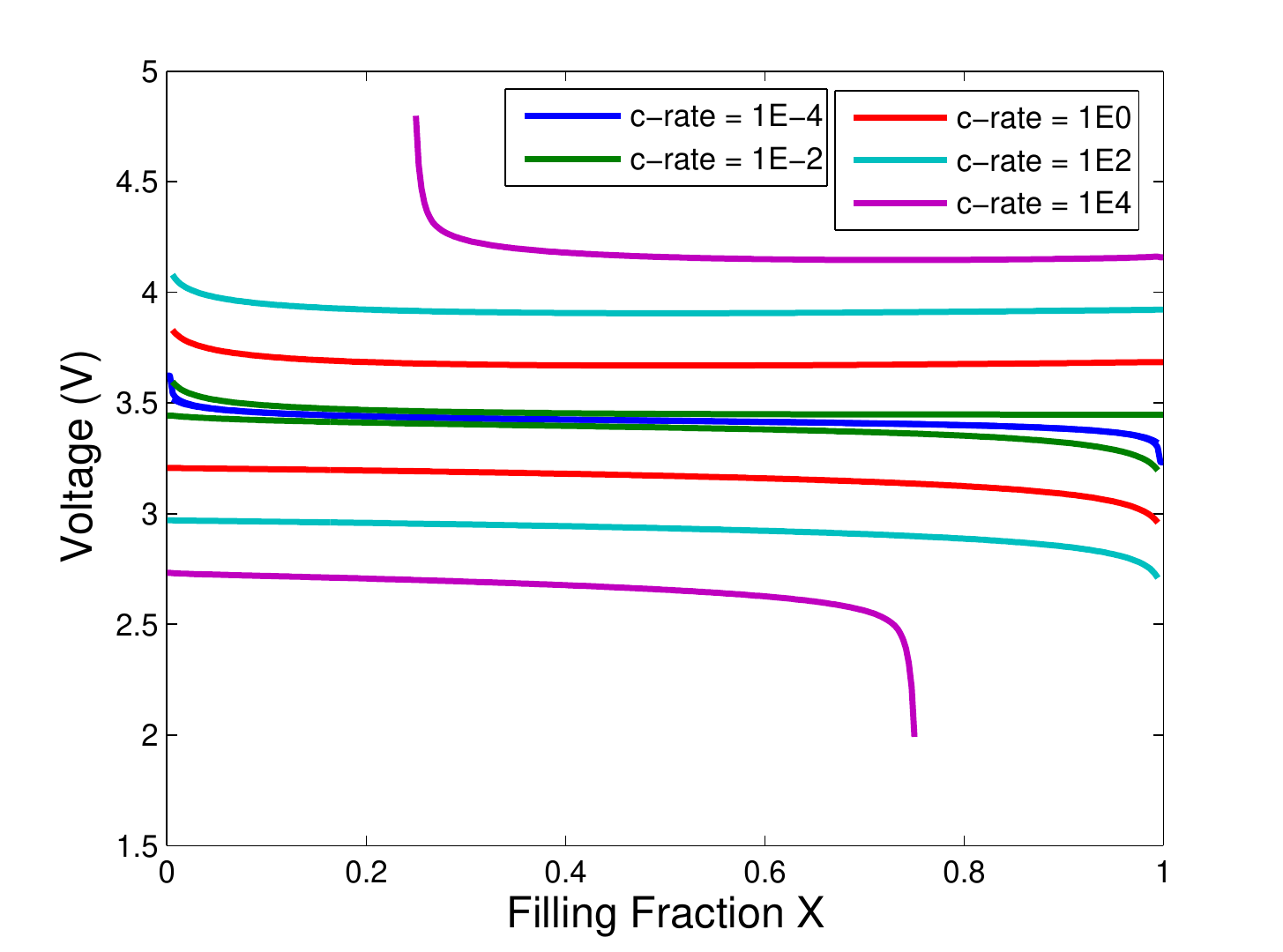}
\end{center}
\caption{Cycling of a high temperature solid solution with attractive forces ($\tilde \Omega = 1$) with other parameters from Fig. \ref{fig:NegativeOmega}.}
\label{fig:MultiCurrentsHighTemp}
\end{figure}

For $T>T_c$, the numerical results are consistent solid solution behavior. For example, we use the same parameters in Table \ref{table:2}, except for the $\Omega=2.57\times10^{-2}$ eV, or $\tilde \Omega =1$, so the absolute temperature is twice the critical value, $T/T_c=2$.  As shown in Fig. \ref{fig:MultiCurrentsHighTemp},  the voltage  varies less strongly with filling fraction, in a way that resembles previous empirical fits of the flat voltage plateau (below) signifying phase separation. There is no phase separation, however, and the concentration profile (not shown) is very similar to the case of repulsive interactions in Fig. \ref{fig:NegativeOmega}. 

\subsection{ Capacity} 

\begin{figure} [!h]
\begin{center}
\includegraphics[scale=0.5]{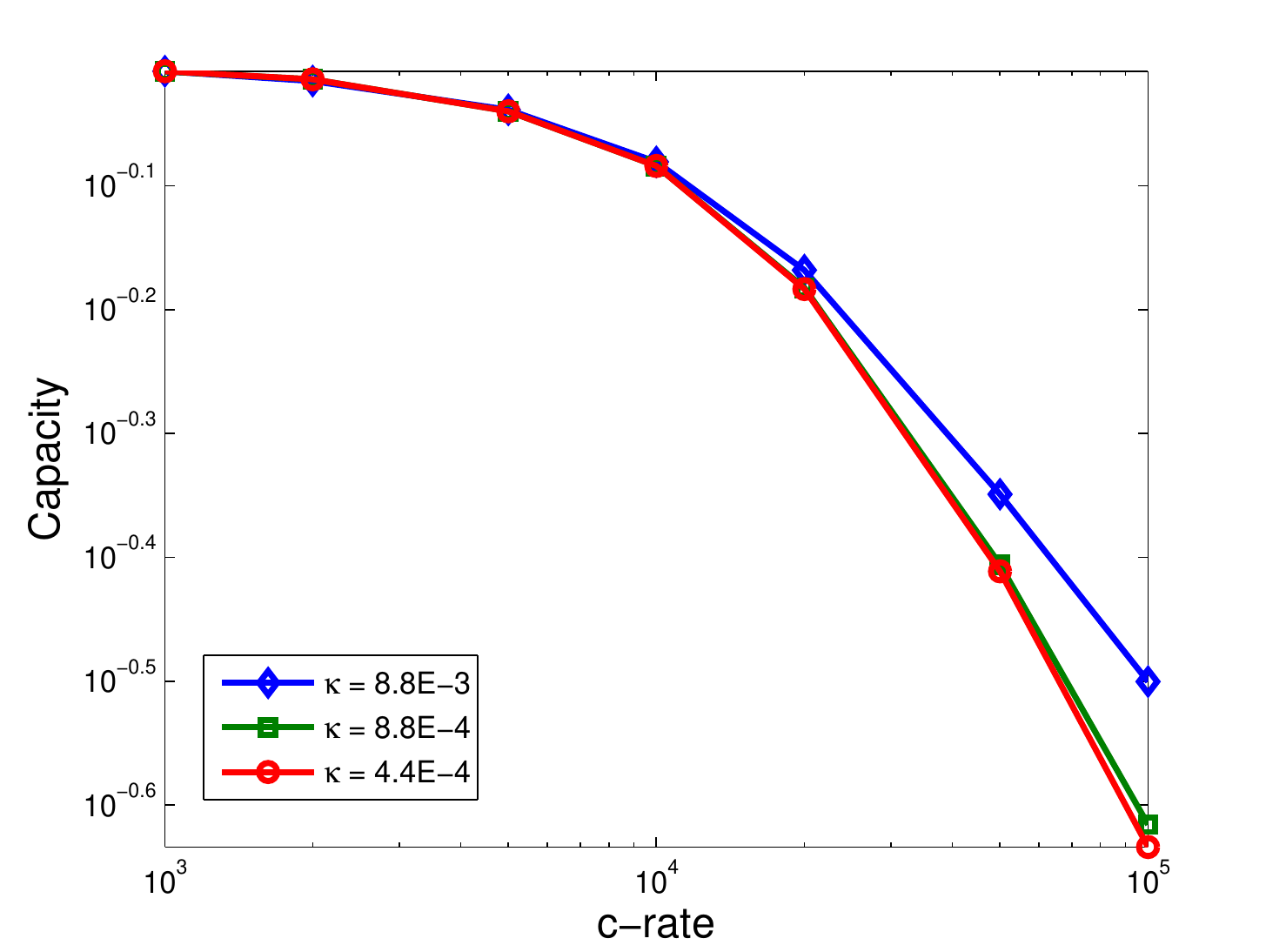}
\end{center}
\caption{Capacity $C$ versus current with different gradient penalty constant \textcolor{black}{$\tilde \kappa$} in a solid solution ($\tilde \Omega=\beta=0$). }
\label{fig:capacity}
\end{figure}

When the particle is charged or discharged at a high rate, the total capacity, defined as the \textcolor{black}{filling fraction} $X$ reached when the voltage drops below some threshold on discharge, will be significantly reduced. In a simple spherical diffusion model, by the scaling of Sand's time $t_s \sim \frac{1}{I^2}$ \cite{bard_book,10.626} and charge conservation, the total capacity $C$ scales as,
$C = I t_s \sim {I}^{-1}$.  In our CHR model, we observe a different scaling of the capacity from the numerical simulations. In a simple power law expression, $C \sim I^{\gamma}$, the exponent $\gamma$ \textcolor{black}{is no longer simply the constant $-1$ as in the spherical diffusion model and generally depends on material properties}, such as wetting parameter $\beta$, gradient penalty constant $\kappa$, and regular solution parameter $\Omega$. A sample of the scaling dependence on current with different  values of $\kappa$ is shown in Fig. \ref{fig:capacity}, where $\gamma \approx 0.5$.

\section{Phase Separation}
\label{sec:phasesep}
In some materials, such as LFP, the attractive forces between intercalated ions are strong enough to drive phase separation into Li-rich and Li-poor solid phases at room temperature, for $T< T_c$, or $\tilde \Omega >2$ in the regular solution model.  Phase separation occurs because the homogeneous chemical potential is no longer a monotonic function of concentration. This has a profound  effect on battery modeling that is predicted from first principles by the CHR model.

\subsection{ Strong attractive forces or low temperature}
\begin{figure}[!h]
\centering
\includegraphics[scale=0.7]{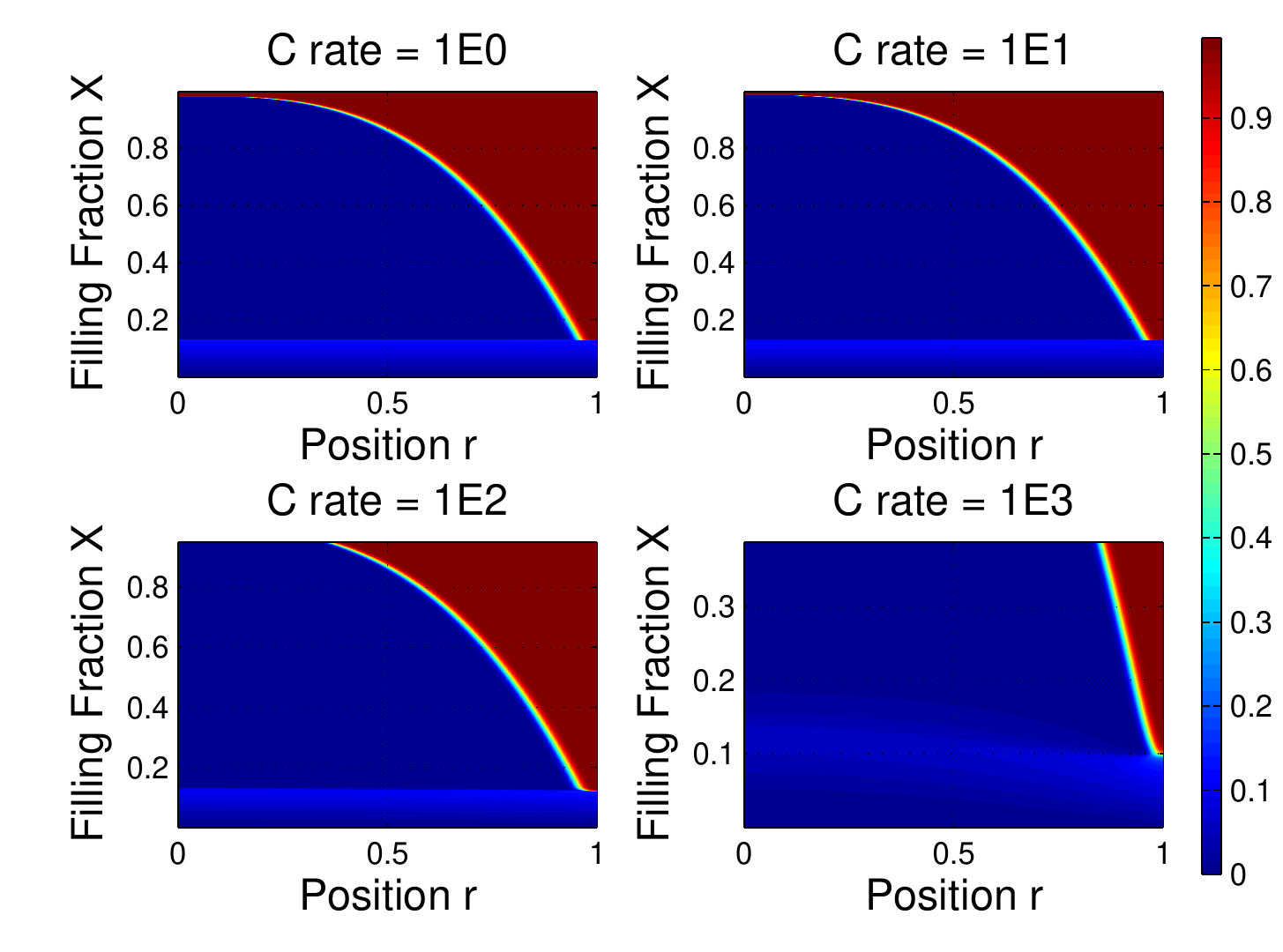}
\caption{Dynamics of phase separation during ion intercalation ($\tilde \Omega=4.48$). Concentration distributions within the spherical particle are shown at different currents \textcolor{black}{for large currents $1C$ (top left), $10C$ (top right), $100C$ (bottom left) and $1000C$ (bottom right).} The x-axis represents the nondimensional radial position $\tilde{r}$ and the y-axis presents the overall average filling fraction $X$ of the whole particle, which can be also seen as the time dimension. The warmer color in the figure indicates a higher local filling fraction.}
\label{fig:concentrationPhaseSeparation}
\end{figure}

\begin{figure} [!h]
\begin{center}
\includegraphics[scale=0.7]{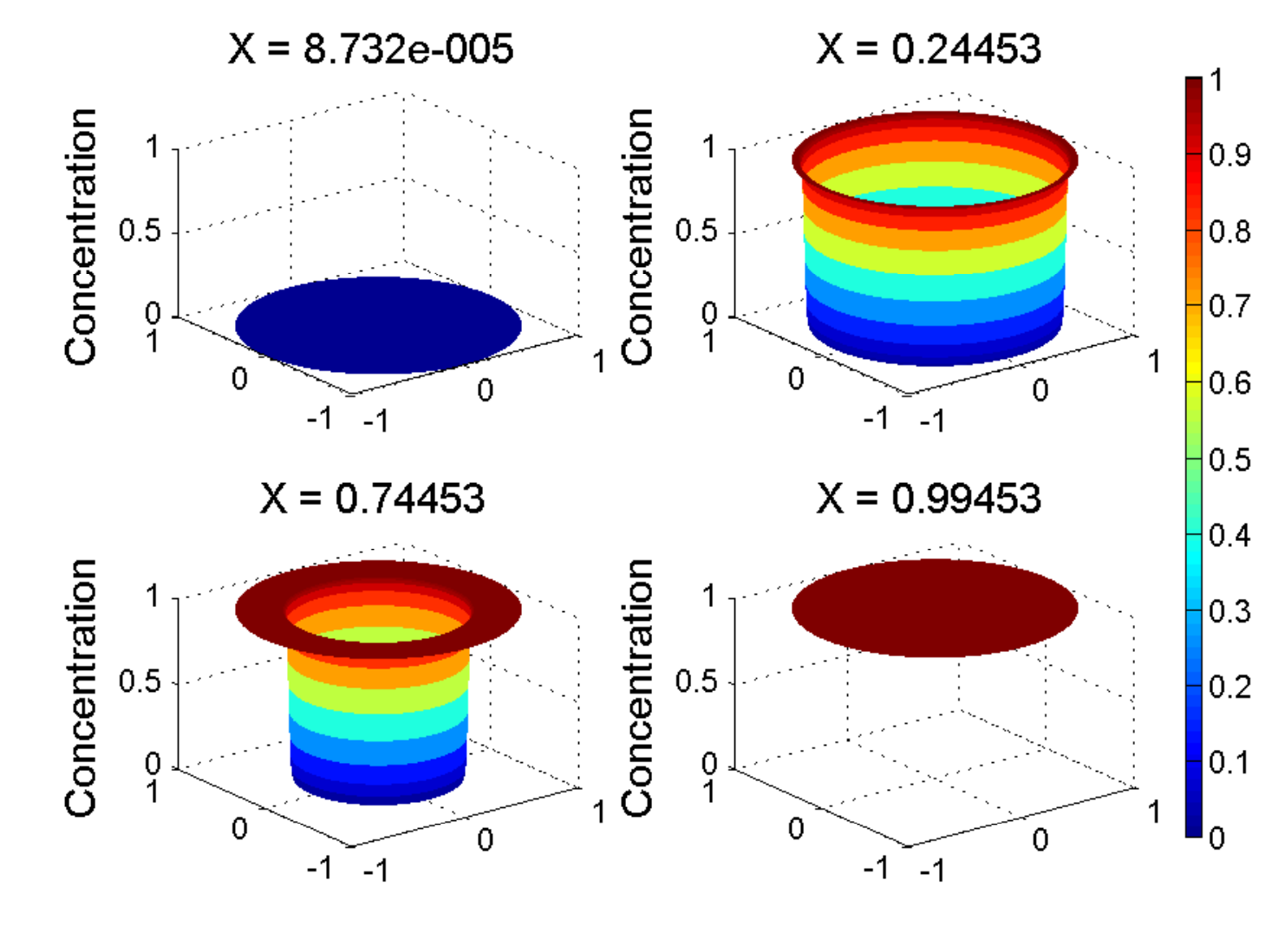}
\end{center}
\caption{ Shrinking core dynamics of phase separation in an isotropic spherical particle ($\tilde \Omega =  4.48$ and no surface wetting). The vertical dimension in the plots shows the concentrations, while the horizontal circle denotes the \textcolor{black}{planar cross section} at the equator of the sphere. \textcolor{black}{The current is of $1C$} and $X$  the overall filling fraction of lithium ions.}
\label{fig:NoWettingConcentration}
\end{figure}

In order to simulate a representative model, we again use the parameters in Table \ref{table:2} but set the $\Omega = 1.15 \times 10^{-1}$ eV, or $\tilde \Omega = 4.48>2$, which is a realistic value of the enthalpy per site value in LFP~\cite{cogswell2012}. Very different from the uniformly filling behavior in Fig. \ref{fig:NegativeOmega}, phase separation occurs suddenly when the composition passes the linearly unstable spinodal points \textcolor{black}{in $\mu$}. The concentration profiles develop sharp  boundaries between regions of uniform composition corresponding to the two stable phases, as shown in Fig. \ref{fig:concentrationPhaseSeparation}.  The new phase appears at the surface and propagates inward, as shown in  Fig. \ref{fig:NoWettingConcentration}, once the surface concentration enters the unstable region of the phase diagram.

After phase separation occurs, the CHR model for an isotropic spherical particle predicts similar two-phase dynamics as the shrinking core model, but without \textcolor{black}{empirically} placing a sharp phase boundary. Instead, the diffuse phase boundary appears from an initial single-phase solid solution at just the right moment, determined by thermodynamic principles, and there is no need to solve a moving boundary problem for a sharp interface, which is numerically simpler.

\begin{figure}[!h]
\centering
\includegraphics[scale=0.4]{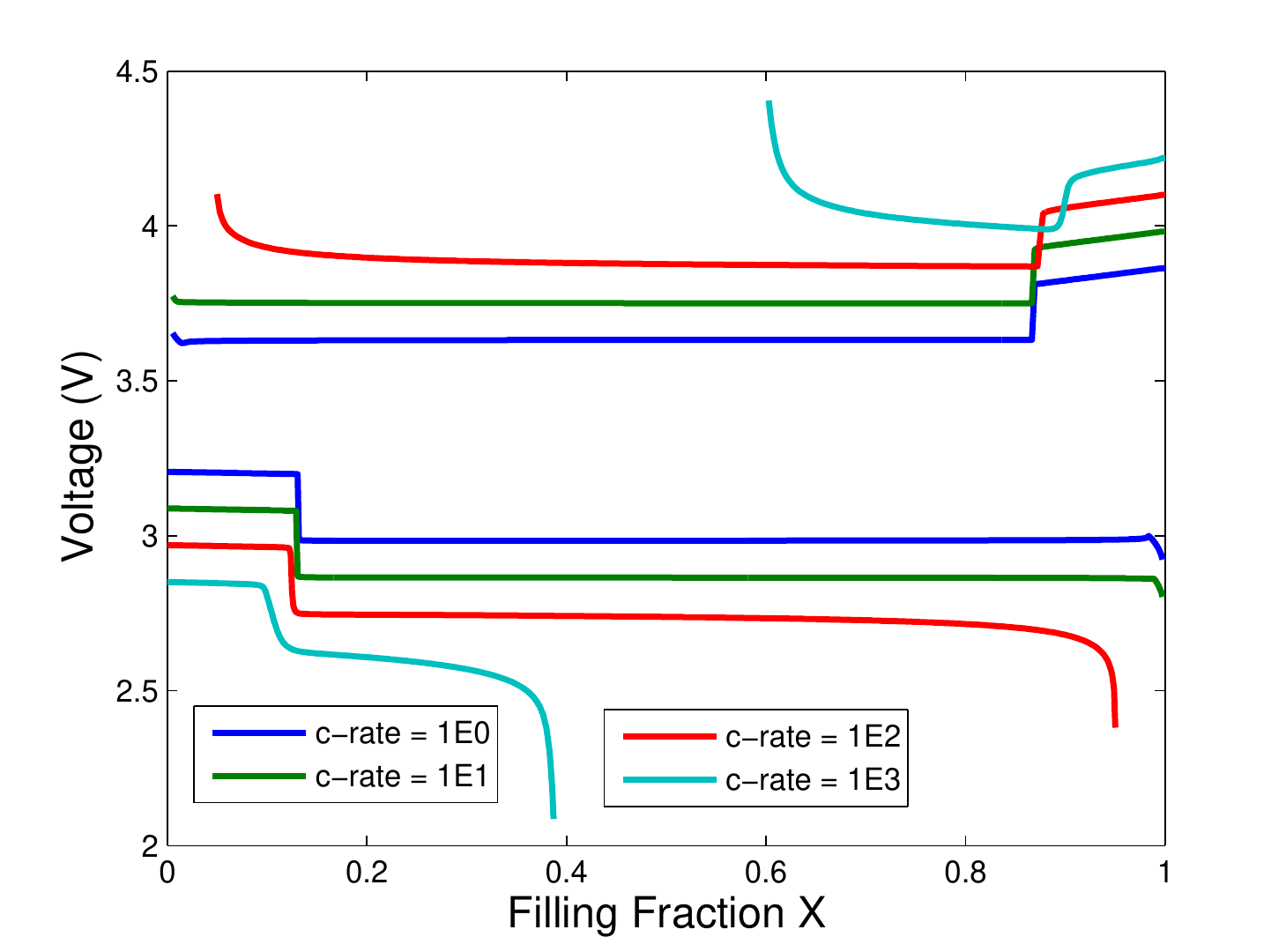}
\caption{Phase separating particle ($\tilde \Omega=4.48$) voltage vs. filling fraction plot at different C-rates \textcolor{black}{ $\pm 1$C, $\pm 10$C, $\pm 100$C and $\pm 1000$C.}}
\label{fig:MultiCurrentsNoWetting}
\end{figure}

The CHR model also predicts the subtle electrochemical signatures of phase separation dynamics~\cite{bazant2013}. Without any empirical fitting,  phase separation naturally leads to a flat voltage plateau,  as shown in Fig. \ref{fig:MultiCurrentsNoWetting}. The constant-voltage plateau reflects the constant chemical potential of ion intercalation in a moving phase boundary (in the absence of coherency strain, which tilts the plateau~\cite{cogswell2012}). At high currents, the initial charge transfer resistance, or activation overpotential, is larger, as signified by the jump to the plateau voltage (derived below), and over time, solid diffusion limitation, or concentration overpotential, causes the voltage to fall more rapidly  during discharging, or increase during charging. 

\subsection{Voltage Plateau Estimation}

\begin{figure} [!h]
\begin{center}
\includegraphics[scale=0.4]{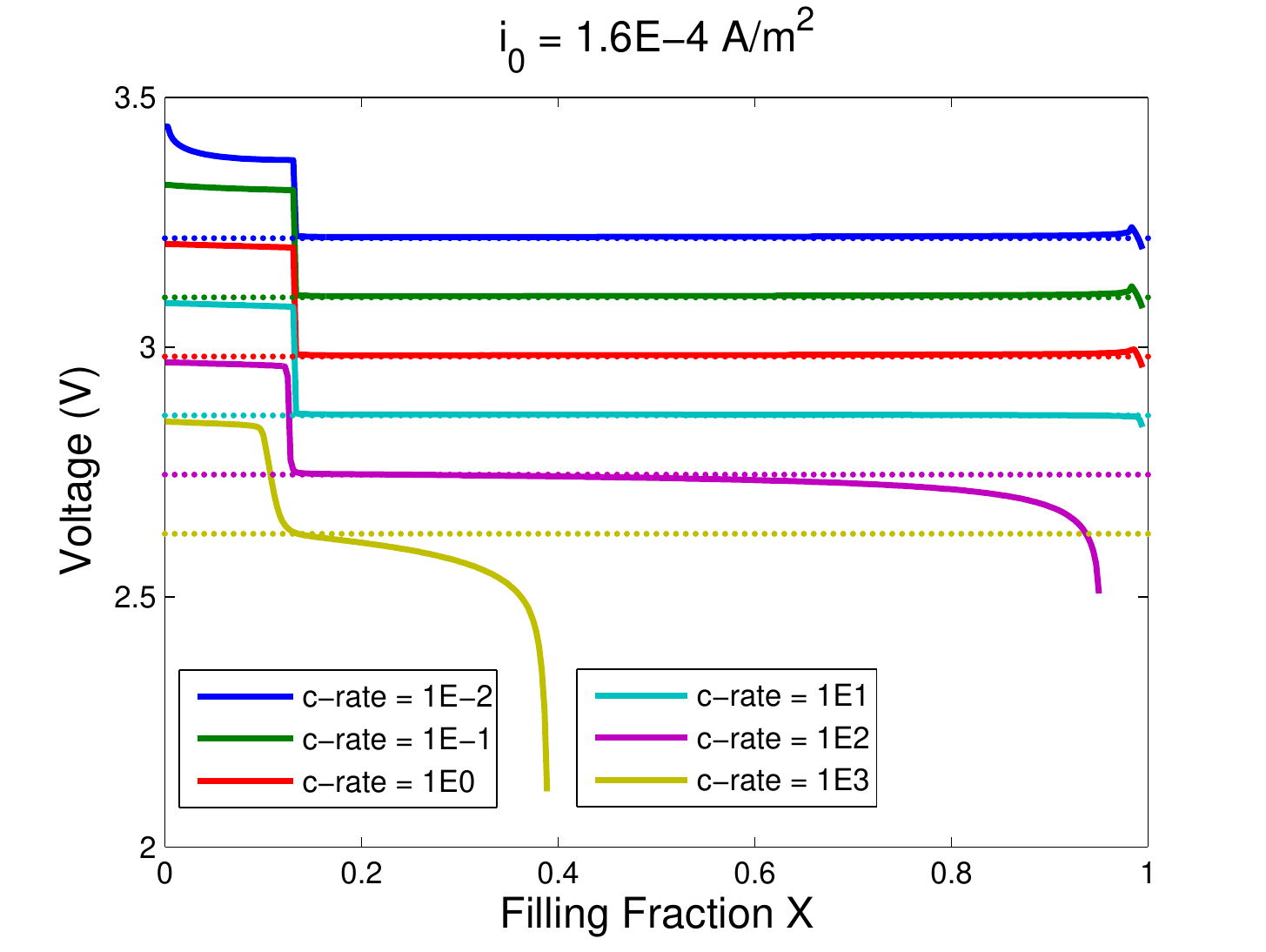}
\end{center}
\caption{ Comparison of the simulated voltage plateau from Fig. \ref{fig:MultiCurrentsNoWetting} (solid curves) and the analytical approximation of Eq. (\ref{eq:Va}) (dashed curves) for $I > 0$. }
\label{fig:PlateauPrediction}
\end{figure}

As we see from Fig. \ref{fig:concentrationPhaseSeparation}-\ref{fig:MultiCurrentsNoWetting}, our model system always undergoes phase separation, which leads to a voltage plateau. In the case without surface wetting, i.e. $\beta = 0$, we can derive an accurate approximation of the voltage plateau \textcolor{black}{value}, since the concentration within each phase is relatively uniform, especially when the current is not very large. Therefore, we may ignore the gradient penalty term $\kappa \nabla^2 c$, leaving only the homogeneous chemical potential, 
\begin{equation}
\tilde \mu \approx \ln\frac{\tilde c}{1 - \tilde c} + \tilde \Omega (1 - 2 \tilde c). 
\end{equation}
The stable composition of each phase approximately solves  $\tilde \mu = 0$, where the homogeneous free energy at these two concentrations takes its minimum. During ion insertion, the surface concentration is approximately the larger solution $\tilde c_l$ of this equation. In the case $I>0$, the plateau voltage is given by 
\begin{equation}
V 
%= V^{\Theta} + \frac{k_BT}{e} (-\tilde \mu - 2 \sinh^{-1} (\frac{\tilde I}{2 \tilde I_0(\tilde c)}) ) 
\approx V^{\Theta} -\frac{2k_BT}{e}\sinh^{-1}\left(\frac{\hat{I}}{4(1-\tilde c_l)}\right).   \label{eq:Va}
\end{equation}
where  $\hat{I}= \frac{\tilde I}{ \tilde {I}_0(\tilde c = \frac{1}{2})}$ is the ratio of the applied current to the exchange current at half filling
At low currents, the agreement between this analytical approximation and the numerically determined voltage plateau is excellent, as shown in \ref{fig:PlateauPrediction}. 

The voltage profile can be understood physically as follows. As a result of our assumption of spherical symmetry, the intercalation reaction must proceed into the outer  ``shell phase", \textcolor{black}{ which is metastable and resists insertion/extraction reactions. This thermodynamic barrier leads to the voltage jumps associated with phase separation in Fig.\ref{fig:PlateauPrediction}. }In the case of lithiation, the shell has high concentration and thus strong entropic constraints inhibiting further insertion that lower the reaction rate, increase the overpotential, and lower the voltage plateau when phase separation occurs. \textcolor{black}{ This behavior is very different from anisotropic models, where the phase boundary is allowed to move along the surface as an intercalation wave~\cite{singh2008,bai2011,cogswell2012,cogswell2013} and insertion occurs with higher exchange current at intermediate concentrations, although the active area is reduced, which leads to suppression of surface phase separation at high currents~\cite{bai2011,cogswell2012}, since higher exchange current density immediately leads to a higher charging/discharging current density, if the ratio of them does not change.}

\subsection{Butler-Volmer Transfer Coefficient}

\begin{figure}[!h]
\begin{center}
\begin{tabular}{ c c}
  \includegraphics[width=0.5 \columnwidth]{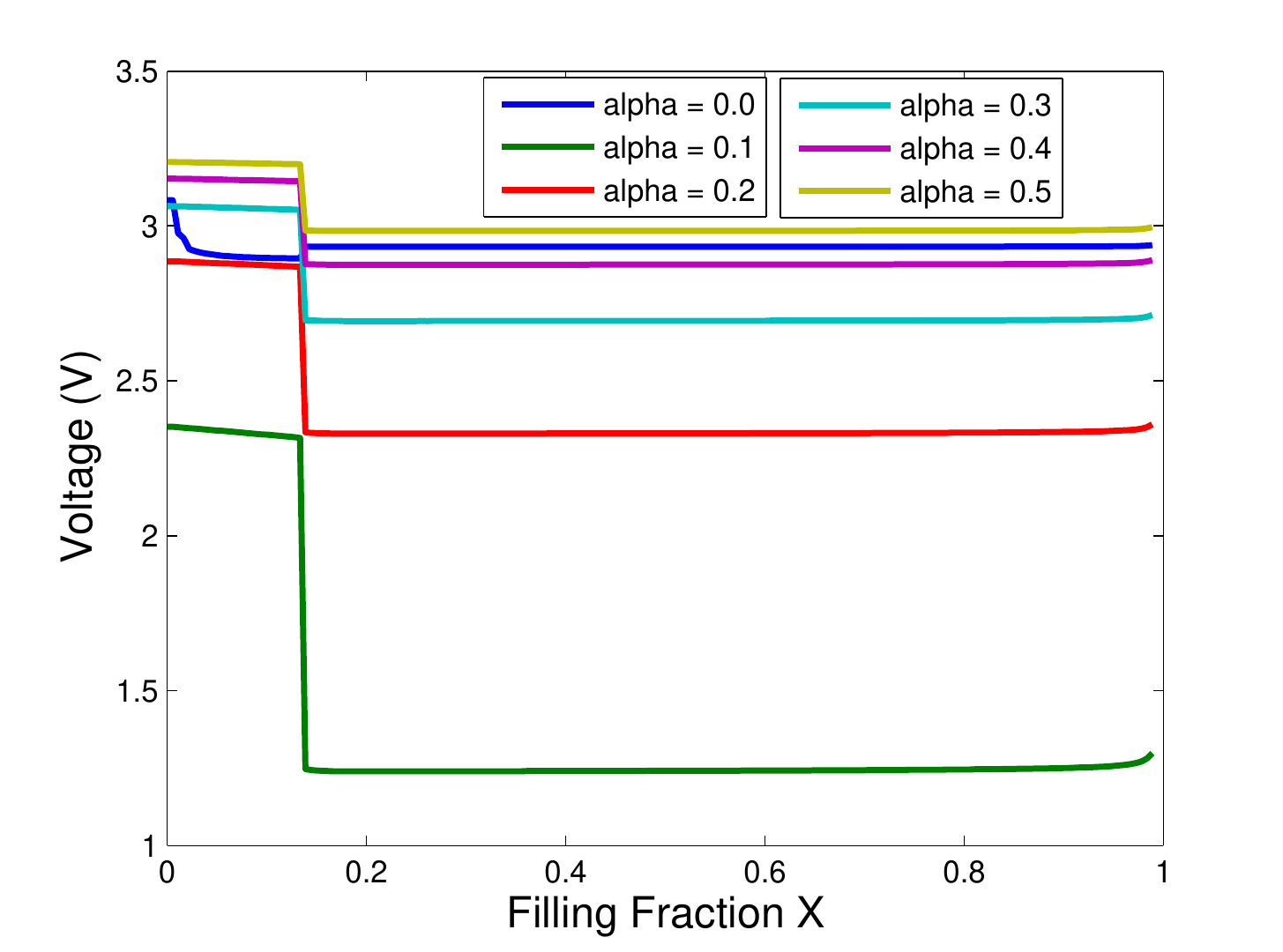} & \includegraphics[width=0.5 \columnwidth]{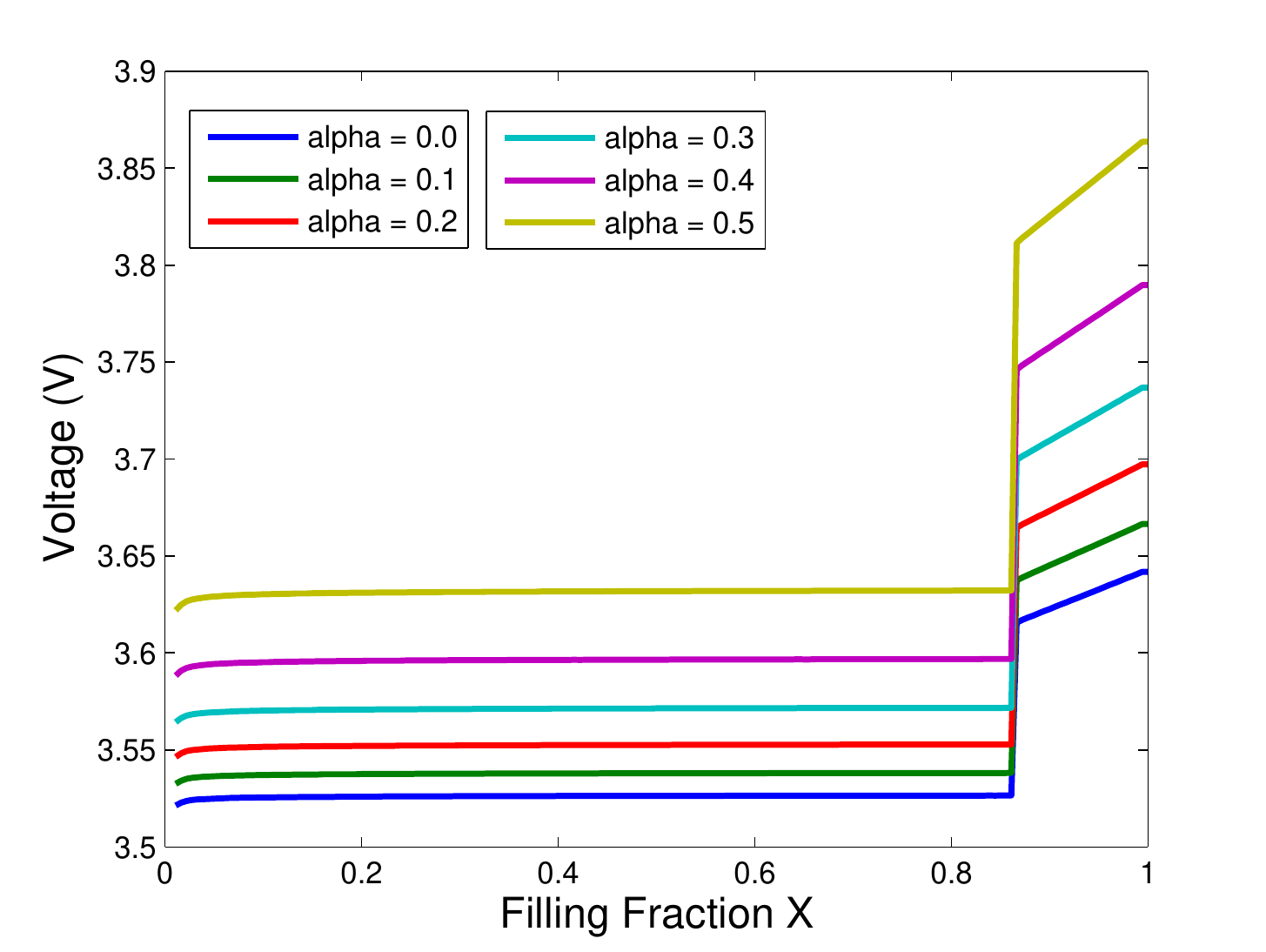} \\   
\end{tabular}
\end{center}
\caption{ Effect of the Butler-Volmer charge transfer symmetry coefficient $\alpha$ on the voltage during battery discharging (left) and charging (right) with \textcolor{black}{constant current $\pm 1C$.} }
\label{fig:MultiAlpha}
\end{figure}

In the preceding examples, we set the Butler-Volmer the transfer coefficient to $\alpha=0.5$ as in prior work with both CHR~\cite{bai2011,cogswell2012} and diffusive~\cite{doyle1993,srinivasan2004} models. This choice can be justified by Marcus theory of charge transfer when the reorganization energy is much larger than the thermal voltage~\cite{bazant2013,bard_book}, but in battery materials this may not always be the case.  In our isotropic model, charge-transfer asymmetry ($\alpha \neq 0.5$) mainly manifests itself via strong broken symmetry between charge and discharge in the activation overpotential, as shown in the voltage plots of  Fig. \ref{fig:MultiAlpha}. A smaller value of $\alpha$ leads to a lower voltage plateau while discharging ($I>0$), but does not much affect the voltage plateau during charging ($I<0$). 

\section{Phase Separation with Surface Wetting}
\label{sec:wet}

The wetting of a solid surface by two immiscible fluids, such as water and air, is very familiar, but it is not widely appreciated that analogous phenomena also occur when binary solids \textcolor{black}{``}wet" a fluid or solid surface and play a major role in nanoparticle intercalation~\cite{cogswell2013}.  The only major difference is that coherent (defect-free) solid-solid interfaces have much lower tension than solid-fluid interfaces due to stretched, rather than broken, bonds. As a result, a stable contact angle cannot form, and  one phase tends to fully wet each surface in equilibrium ($\Theta_c=0,\pi$), regardless of the bulk composition.   The competition between different phases to wet a surface can promote the nucleation of a phase transformation via the instability of a surface wetting layer.   In particular, the wetting of certain crystal facets of LFP particles by either LiFePO$_4$ and FePO$_4$ ensures the existence of surface layers that can become unstable and propagate into the bulk, as a means of surface-assisted nucleation~\cite{cogswell2013}.

\subsection{ Shrinking cores and expanding shells} 
In this section, we show that surface wetting characteristics have a significant effect on the concentration profile and voltage during insertion, even in an isotropic spherical particle.  
Mathematically, we impose the inhomogeneous Neumann boundary condition, $\frac{\partial \tilde{c}}{\partial \tilde r}(1,\tilde{t}) = \beta$, where, as described above, $\beta > 0$ promotes the accumulation of ions at the surface, or wetting by the high density phase.  
In this case, during ion insertion,   the surface concentration will be always higher than the remaining bulk region, if we start from a uniform low concentration. As a result, the surface hits the spinodal point earlier than other places inside the particle, which means the Li-rich phase always nucleates at the surface. In an isotropic particle, this leads to the shrinking core phenomenon, as in the cases without surface wetting ($\beta=0$) described above.  

The case of surface de-wetting  ($\beta<0$) is interesting because surface nucleation is suppressed, and more than two phase regions can appear inside the particle. During insertion, the surface concentration is now always lower than in the interior, especially when the current is small. Therefore, an interior point will reach the spinodal concentration earlier than the surface, so the high-density phase effectively nucleates somewhere in the bulk, away from the surface.  

\begin{figure}[!h]
\centering
 \includegraphics[scale=0.5]{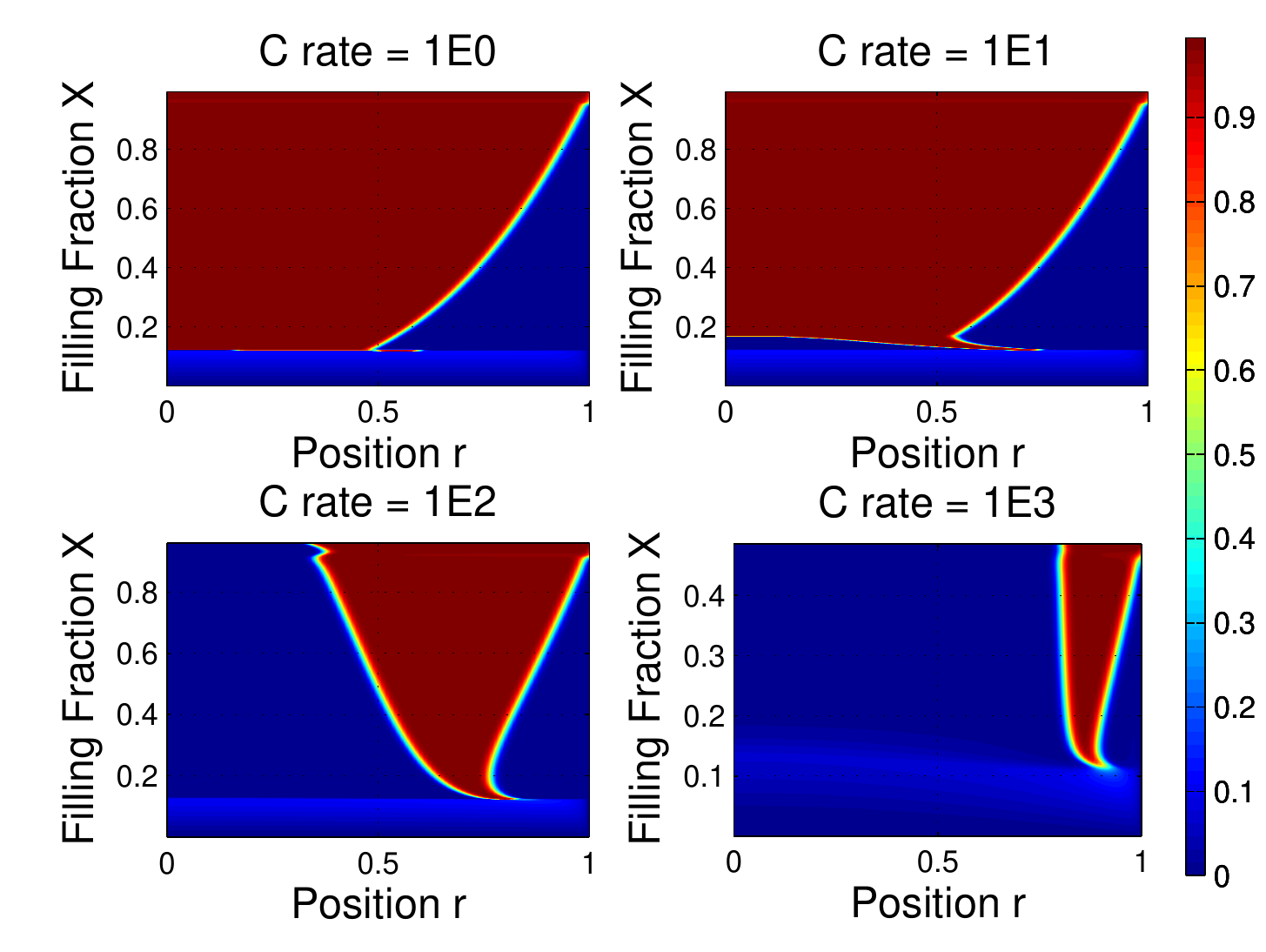} 
\caption{ Phase boundary motion during ion insertion in  a spherical particle with surface de-wetting ($\beta = -17.9$, $\Omega=4.48$)  at different large currents \textcolor{black}{ $1$C, $10$C, $100$C and $1000$C.} The warmer color in the figure indicates a higher local filling fraction.}
\label{fig:ConcentrationDewetting}
\end{figure}

\begin{figure} [!h]
\begin{center}
\begin{tabular}{ c c}
\includegraphics[scale=0.4]{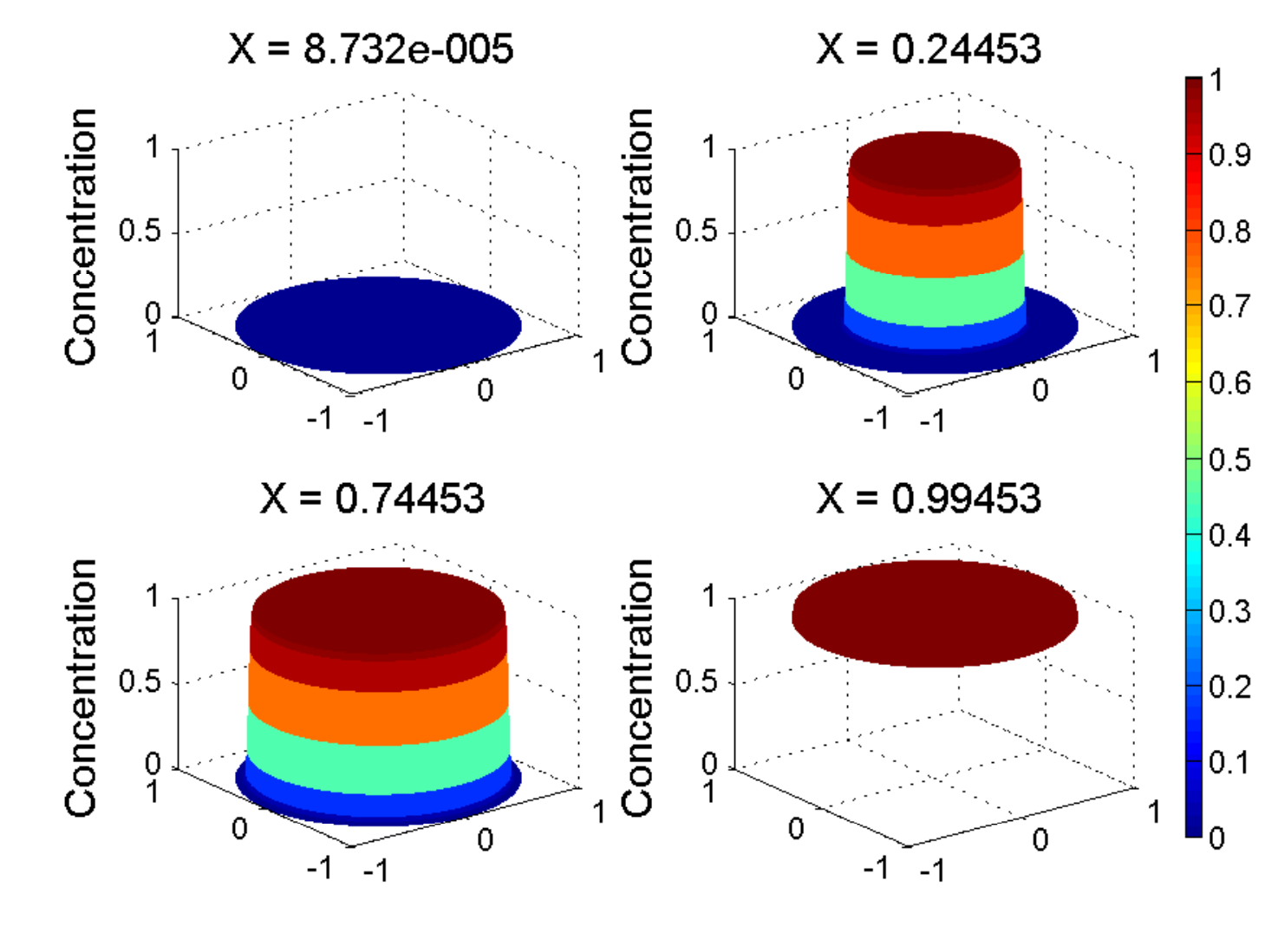} & \includegraphics[scale=0.4]{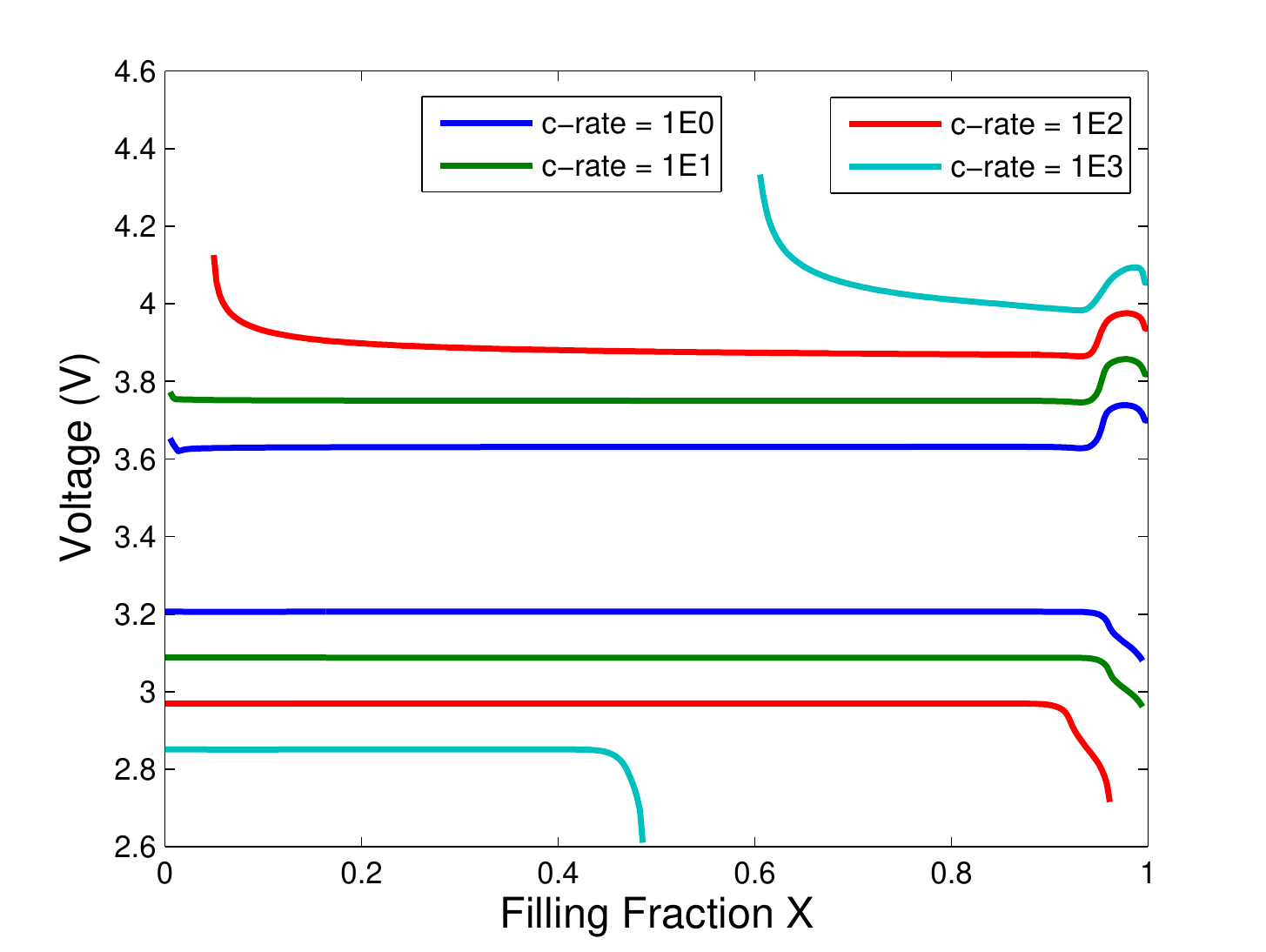}\\   
\end{tabular}
\end{center}
\caption{ Concentration profiles (left) and voltage transients (right) for ion insertion at currents \textcolor{black}{ $\pm 1$C, $\pm 10$C, $\pm 100$C and $\pm 1000$C} in a phase separating spherical particle ($\tilde \Omega =  4.48$ and surface de-wetting $\beta= -17.9$).}
\label{fig:Dewetting}
\end{figure}

As a result, there is an ``expanding shell" at the same time as a shrinking core of the low density phase.  
This unusual behavior is shown in Fig. \ref{fig:ConcentrationDewetting} for $\beta= -17.9$ at several currents. The surface energy is $\gamma = -90$ mJ/m$^2$ at maximum filing, if we assume the $\gamma$ is a linear function of concentration. A detailed demonstration of this concentration dynamics is shown in  Fig. \ref{fig:Dewetting}.  The middle Li-rich region expands inward and outward simultaneously, it first fills up the Li-poor phase located at the center, and finally it fills the whole particle.

\begin{figure} [!h]
\begin{center}
\begin{tabular}{ c c}
\includegraphics[width=0.5 \columnwidth]{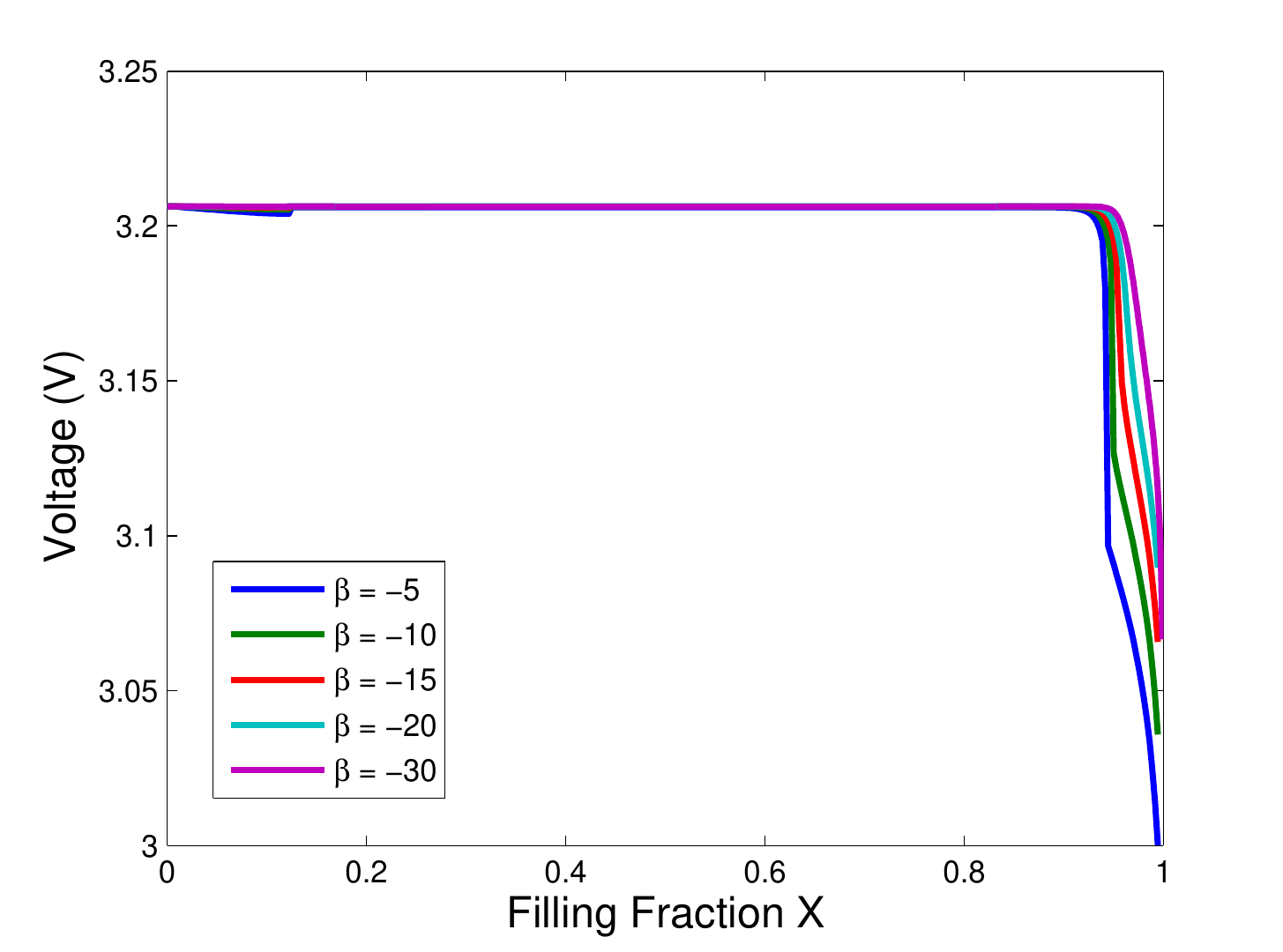} & \includegraphics[width=0.5 \columnwidth]{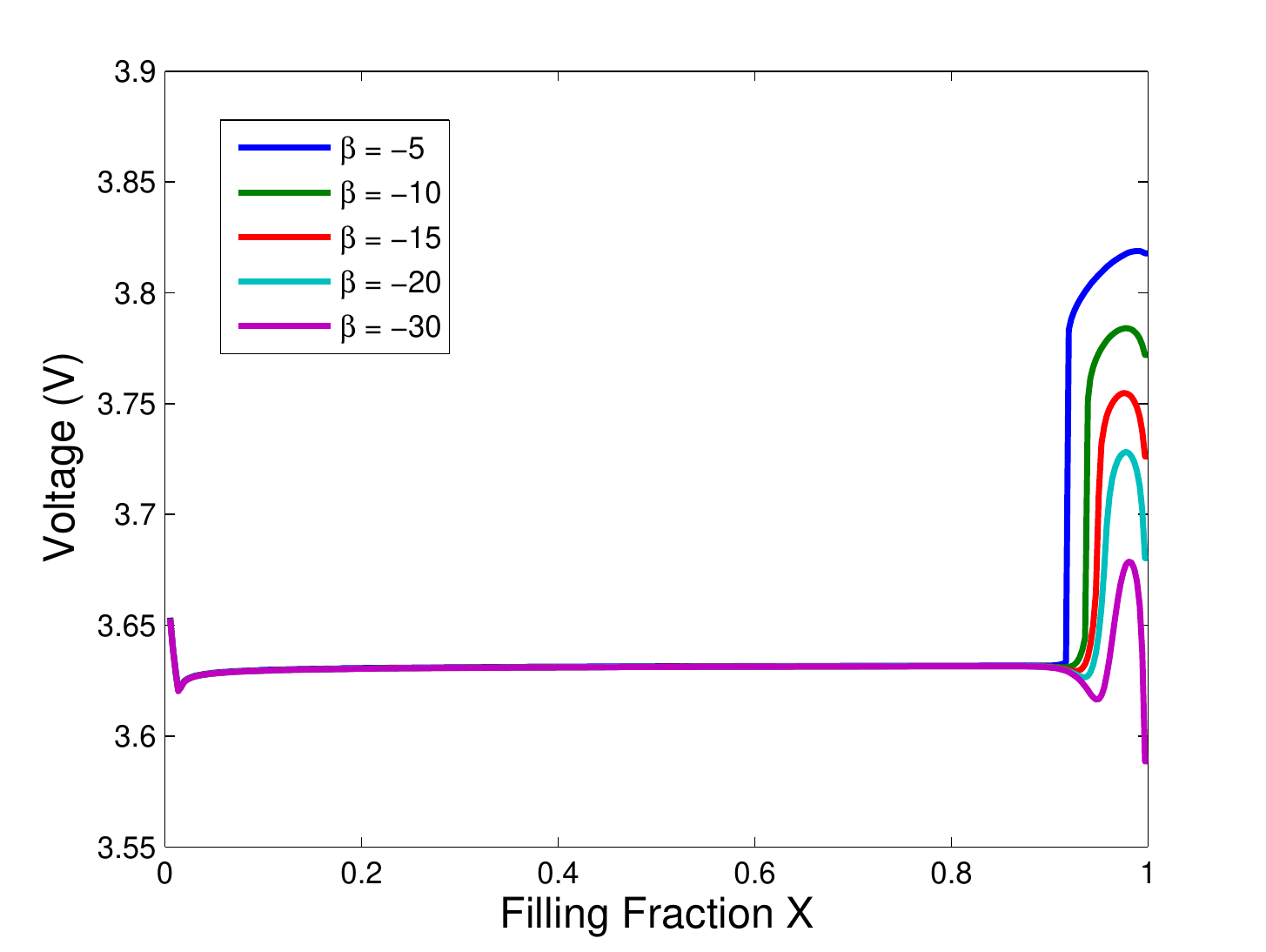}
\end{tabular}
\end{center}
\caption{Effect of a negative  surface wetting parameter ($\beta<0$)  on the voltage during discharging at \textcolor{black}{ $1$C (left) and charging 
at $-1$C.}  }
\label{fig:MultiDerivatives}
\end{figure}

Since the surface is always in the lower stable concentration after the initial phase separation, which does not vary according to the surface derivative $\beta$, we should expect the voltage has very weak dependence on the surface de-wetting condition. The voltage - filling fraction plot in Fig. \ref{fig:MultiDerivatives} confirms this intuition. When $I<0$, the strong surface de-wetting will make the surface concentration very closed to zero, which will make the chemical potential extremely sensitive to small perturbation in concentration, therefore, we only show the results with relatively weak surface de-wetting ($\beta \geq -10$).

\subsection{Voltage efficiency}

In the limit of zero current at a given filling, the voltage given by the Nernst equation has a unique value $V(X)$ corresponding to thermodynamic equilibrium. When a current is applied, energy is lost as heat due to various resistances in the cell, and there is a voltage gap $\Delta V$ between charge and discharge at the same filling. The voltage  efficiency is $1 - \Delta V / V_0$. To account for transient effects, we define the voltage gap for a given current magnitude $|I|$ as the voltage at half filling ($X=0.5$) during galvanostatic charging starting from nearly full with $I<0$, minus that during discharging starting from nearly empty with $I>0$.

\begin{figure} [!h]
\begin{center}
\includegraphics[scale=0.5]{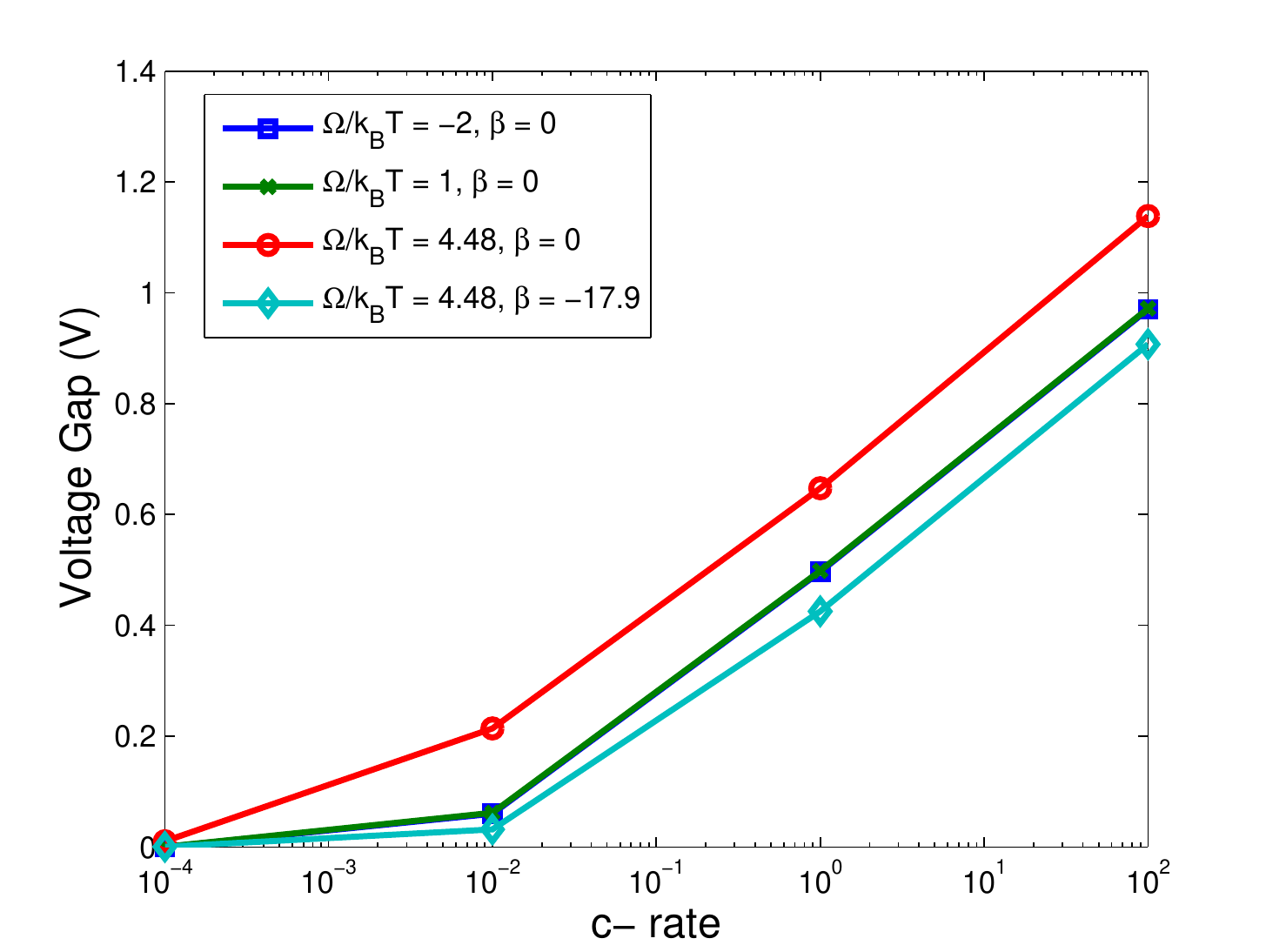}
\end{center}
\caption{The gap of the charging and the discharging voltage when the particle is half filled, $X = 0.5$, under several conditions including current, $\tilde \Omega$ and surface wetting. The $\beta$ shown in the legend is the nondimensional concentration derivative at the particle surface, which denotes the surface wetting condition.}
\label{fig:VoltageGap}
\end{figure}

In Fig. \ref{fig:VoltageGap}, we show how different parameters, such as the current, mixing enthalpy, and surface wetting condition affect the voltage gap.  For our single particle model with surface nucleation, the voltage gap vanishes at zero current, in contrast to experiments~\cite{dreyer2010} and simulations~\cite{dreyer2011,ferguson2012,ferguson2014,orvananos2014} with porous multi-particle electrodes.  There is  no contradiction, however, because the zero-current voltage gap is an emergent property of a collection of particles with two stable states, resulting from the mosaic instability of discrete transformations (which can also been seen in an array of balloons~\cite{dreyer2011}).

In the case without surface wetting (\textcolor{black}{$\beta=0$}),  the voltage gap is smaller for solid solutions ($\tilde \Omega < 2$) than for phase separating systems ($\tilde \Omega > 2$), since it is more difficult to insert ions into the \textcolor{black}{high concentration} stable state than into an intermediate concentration.  With strong surface de-wetting by the ions ($\beta < 0$) and phase separation ($\tilde \Omega > 2$), however, the gap can be even smaller than in the solid solution case without surface wetting, because the persistence of the low density phase promotes easy intercalation.  This is an important observation because it shows the possibility of improving the voltage efficiency by engineering the solid-solid contact angle of the active particles.

\section{Numerical Methods and Error Convergence}
\label{sec:num}

The CHR model is fourth-order in space and highly nonlinear and thus requires care to solve numerically with accuracy and efficiency.  Naive finite difference or finite volume methods would be unstable or inaccurate.  In order to obtain the solutions above, we developed a new conservative numerical scheme to solve the CHR model with second-order accurate discretization, described in this section.

\subsection{Numerical Scheme}

Great effort has been devoted for solving the Cahn-Hilliard equation numerically with different boundary conditions, and several numerical schemes have been employed, e.g. finite difference \cite{choo1998,de2005,shin2011}, finite element \cite{Banas2008, zhang2010, wodo2011}, spectral method \cite{he2009}, boundary integral \cite{dehghan2009}, level set \cite{greer2006}, discontinuous Galerkin \cite{xia2007} and multi-grid methods \cite{kim2004,wise2007}. 

As our problem is associated with the flux boundary condition, the finite volume method is a more convenient and suitable choice for discretization \cite{burch_thesis,cueto2008,dargaville2013numerical}. Furthermore, the finite volume method may be superior to other methods by its perfect mass conservation and the capability for capturing the concentration shock during phase separations. 

The finite volume method handles the integral form of the Cahn-Hilliard equation. Using the divergence theorem we may update the change of average concentration within a small volume by calculating the difference of \textcolor{black}{the inward and outward fluxes} over the corresponding volume boundary. In the recent literature, two basic approaches for estimating the concentrations and their derivatives at the boundary have been developed. 

Burch \cite{burch_thesis} uses the finite difference type technique to extrapolate the desired unknown values with the known average concentration in each control volume. This approximation method is highly efficient in low dimensional cases with a well-structured grid. Cueto-Felgueroso and Peraire \cite{cueto2008}, Dargaville and Farrell \cite{dargaville2013numerical} develop a different least squares based technique, which is more suitable for high dimensions cases with un-structured \textcolor{black}{grids}. They use the concentrations and their partial derivatives on the control volume boundaries to predict the centroid concentrations nearby, and find the ``most probable" boundary values (concentrations and derivatives) by least square minimizing the prediction errors in centroid concentrations.

\textcolor{black}{However, as we are mostly focusing on the activities exactly on the particle surface, the finite volume method can only provide us information about the average concentration in the shell closed to the surface.} It may take additional computation cost to extrapolate the surface condition and this will introduce additional error as well.

In order to avoid such extrapolation, we propose a numerical scheme that can immediately provide information on the particle surface and still keep the benefits of the finite volume method in conservation and shock toleration, which is inspired by our  numerical method for solving the 1D nonlinear spherical diffusion problem \cite{zeng2013numerical}. Similar to the finite volume method, our numerical scheme indeed handles the integral form of the original PDE system.  We work with dimensionless variables, but drop the tilde accents for ease of notation.  Since the phase boundary may propagate to any location in the sphere, a non-uniform mesh may not be as helpful as the case in normal nonlinear diffusion problem, so we use uniform grids.

Consider a  $N$-point uniform \textcolor{black}{radial spatial} mesh within the sphere, $r_1$, $r_2$, $r_3$, $\cdots$, $r_N$, while $r_1=0$ is \textcolor{black}{at} the sphere center and $r_N$ is right on the surface. Here we define that $\Delta r = r_{j+1}-r_j$, for any $j \in \{ 1, 2, \cdots, N-1\}$ and make $c_1$, $c_2$, $c_3$, $\cdots$, $c_N$ to be the concentration on these grid points. 

If we integrate the Eqn. \ref{eqn:MassConservation} over a shell centered at a non-boundary grid point $r_i$ with width $\Delta r$, which is equivalent to the volume $V_i$ between $[r_i-\frac{\Delta r}{2},r_i+\frac{\Delta r}{2} ]$, by divergence theorem we have,
\begin{equation}
\int_{V_i}\frac{\partial c}{\partial t} dV = -\int_{V_i} \nabla \cdot F dV=-\int_{\partial V_i} n \cdot F dS. 
\end{equation}
We can further write both sides of the above equation in the following form,
\begin{equation}
\label{eqn:IntigratedForm}
\int_{r_i-\frac{\Delta r}{2}}^{r_i+\frac{\Delta r}{2}}  4 \pi r^2 \frac{\partial c}{\partial t} dr = 4 \pi ( (r_i-\frac{\Delta r}{2})^2 F_{i-\frac{1}{2}}- (r_i+\frac{\Delta r}{2})^2 F_{i+\frac{1}{2}}).
\end{equation}
while $F_{i-\frac{1}{2}} = F \Big |_{r_i-\frac{\Delta r}{2}}$ and $F_{i+\frac{1}{2}} = F \Big |_{r_i+\frac{\Delta r}{2}}$.

The left hand side of the above Eqn. \ref{eqn:IntigratedForm} can be approximated by,
\begin{equation}
\int_{r_i-\frac{\Delta r}{2}}^{r_i+\frac{\Delta r}{2}}  4 \pi r^2 \frac{\partial c}{\partial t} dr = \frac{\partial }{\partial t} (\frac{1}{8}V_{i-1} c_{i-1} + \frac{3}{4} V_{i}c_i + \frac{1}{8}V_{i+1} c_{i+1} + O(\Delta r^3)).
\end{equation}
This can be also written in a matrix form for each small volume on each row,
\begin{equation}
\left(
\begin{array}{c}
\int_{r_1}^{r_1+\frac{\Delta r}{2}}  4 \pi r^2 \frac{\partial c}{\partial t} dr \\
\int_{r_2-\frac{\Delta r}{2}}^{r_2+\frac{\Delta r}{2}}  4 \pi r^2 \frac{\partial c}{\partial t} dr\\
\int_{r_3-\frac{\Delta r}{2}}^{r_3+\frac{\Delta r}{2}}  4 \pi r^2 \frac{\partial c}{\partial t} dr\\
\vdots\\
\int_{r_{N-1}-\frac{\Delta r}{2}}^{r_{N-1}+\frac{\Delta r}{2}}  4 \pi r^2 \frac{\partial c}{\partial t} dr\\
\int_{r_N-\frac{\Delta r}{2}}^{r_N}  4 \pi r^2 \frac{\partial c}{\partial t} dr
\end{array}
\right) \approx
\textbf{M} \frac{\partial }{\partial t}
\left(
\begin{array}{c}
c_1\\
c_2\\
c_3\\
\vdots\\
c_{N-1}\\
c_N.
\end{array}
\right),
\end{equation}
while $\textbf{M}$ is the mass matrix,
\begin{equation}
\textbf{M}=
\left(
\begin{array}{cccccccc}
\frac{3}{4} V_1 & \frac{1}{8} V_2 & 0 & 0 & \cdots & 0 & 0 & 0\\
\frac{1}{4} V_1 & \frac{3}{4} V_2 & \frac{1}{8} V_3 & 0 & \cdots  & 0 & 0 & 0\\
0 & \frac{1}{8} V_2 & \frac{3}{4} V_3 & \frac{1}{8} V_4 & \cdots  & 0 & 0 & 0 \\
\vdots & \vdots & \vdots & \vdots & \ddots  & \vdots & \vdots & \vdots\\
0 & 0 & 0 & 0 & \cdots & \frac{1}{8} V_{N-2} & \frac{3}{4} V_{N-1} & \frac{1}{4} V_N\\
0 & 0 & 0 & 0 & \cdots & 0 & \frac{1}{8} V_{N-1} & \frac{3}{4} V_N\\
\end{array}
\right).
\end{equation}

In fact, this is the major \textcolor{black}{improvement} of our method from the classical finite difference method. Instead of having a diagonal mass matrix in the finite volume method, we hereby use a tri-diagonal mass matrix in our new numerical scheme. Since each column of the this matrix sum to the volume of the corresponding shell, this indicates \textcolor{black}{that} our method must conserve mass with a correct volume.

Before we approximate the flux $F$, we will give the approximation formula for the chemical potential $\mu_i$ at each grid point $r_i$. when $i = 2$, $3$, $\cdots$, $N-1$, 
\begin{eqnarray}
\begin{split}
\mu_i = \ln \frac{c_i}{1 - c_i} + \Omega (1-2c_i) - \kappa \nabla^2 c_i = \ln \frac{c_i}{1 - c_i} + \Omega (1-2c_i) - \kappa(\frac{2}{r_i} \frac{\partial c}{\partial r} + \frac{\partial^2 c}{\partial r^2})\\
 =\ln \frac{c_i}{1-c_i} + \Omega (1-2c_i) - \kappa (\frac{c_{i-1} - 2r_i + c_{i+1}}{\Delta r^2} + \frac{2}{r_i} \frac{c_{i+1} - c_{i-1}}{2 \Delta r}) + O(\Delta r^2).
\end{split}
\end{eqnarray}

For $i=1$, by symmetric condition at the center and the isotropic condition, $\nabla^2 c_1 = 3\frac{\partial^2 c_1}{ \partial r^2}$ and $\nabla c_1 = 0$, then,
\begin{equation}
\mu_1 = \ln \frac{c_1}{1-c_1} + \Omega (1-2c_1) -3\kappa\frac{\partial^2 c_1}{ \partial r^2} = \ln \frac{c_1}{1-c_1} + \Omega (1-2c_1) -3\kappa \frac{2c_2 -2c_1}{\Delta r^2} + O(\Delta r^2).
\end{equation}

For $i=N$, since we have the boundary condition $ n \cdot \kappa \nabla c_N = \frac{\partial \gamma_s}{\partial c}$, when $\frac{\partial \gamma_s}{\partial c}$ is only a constant or a function of $c_N$, we can assume a ghost grid point at $r_{N+1}$, while the concentration at this point satisfies $\nabla c_N = \frac{c_{N+1} - c_{N-1}}{2 \Delta r}=\beta$, which is equivalent to $c_{N+1} = 2 \Delta r \beta + C_{N-1}$, 
\begin{equation}
\mu_N = \ln \frac{c_N}{1-c_N} + \Omega (1-2c_N) - \kappa(\frac{2}{r_N}\beta + \frac{2C_{N-1} -2C_N + 2\Delta r \beta}{\Delta r^2} + O(\Delta r^2)).
\end{equation}

With the chemical potential on each grid point, we can estimate the right hand side of the Eqn. \ref{eqn:IntigratedForm}. For each midpoint of two grid points, the flux $F_{i+\frac{1}{2}}$ satisfies,
\begin{equation}
F_{i+\frac{1}{2}} = -(1-\frac{c_i + c_{i+1}}{2})\frac{c_i + c_{i+1}}{2}\frac{\mu_{i+1} - \mu_i}{\Delta r} + O(\Delta r^2).
\end{equation}

For center of the sphere, again by the symmetric condition we have
\begin{equation}
F \Big |_{r=0} = 0.
\end{equation}
 
And finally for the particle surface the flux is given by the current, which is also our boundary condition.
\begin{equation}
F \Big |_{r=1} = -F_s.
\end{equation}

This finishes the discretization of the original partial differential equations system to a time dependent ordinary differential equations system. We use the implicit $ode15s$ solver for the time integration to get the numerical solution.

\subsection{Error Convergence Order}

As we demonstrated in the derivation of this numerical method, the discretization has \textcolor{black}{error terms in second or higher orders}. Thus, we may expect the error convergence order in the spatial meshing should be also in the second order. This will be confirmed by the numerical convergence test.

In the error convergence test, we use small current density \textcolor{black}{ $10^{-4}$C.} We will also assume no surface wetting in this test. As we are mostly interested in the voltage prediction from this single particle ion-intercalation model, we will define the error as the $L^2$ norm of the difference in voltage comparing to the standard curve, which will use the solution from very fine grid ($3001$ uniform grid points in our case) as the reference solution. 

\begin{figure} [!h]
\begin{center}
\begin{tabular}{ c c}
\includegraphics[width=0.5 \columnwidth]{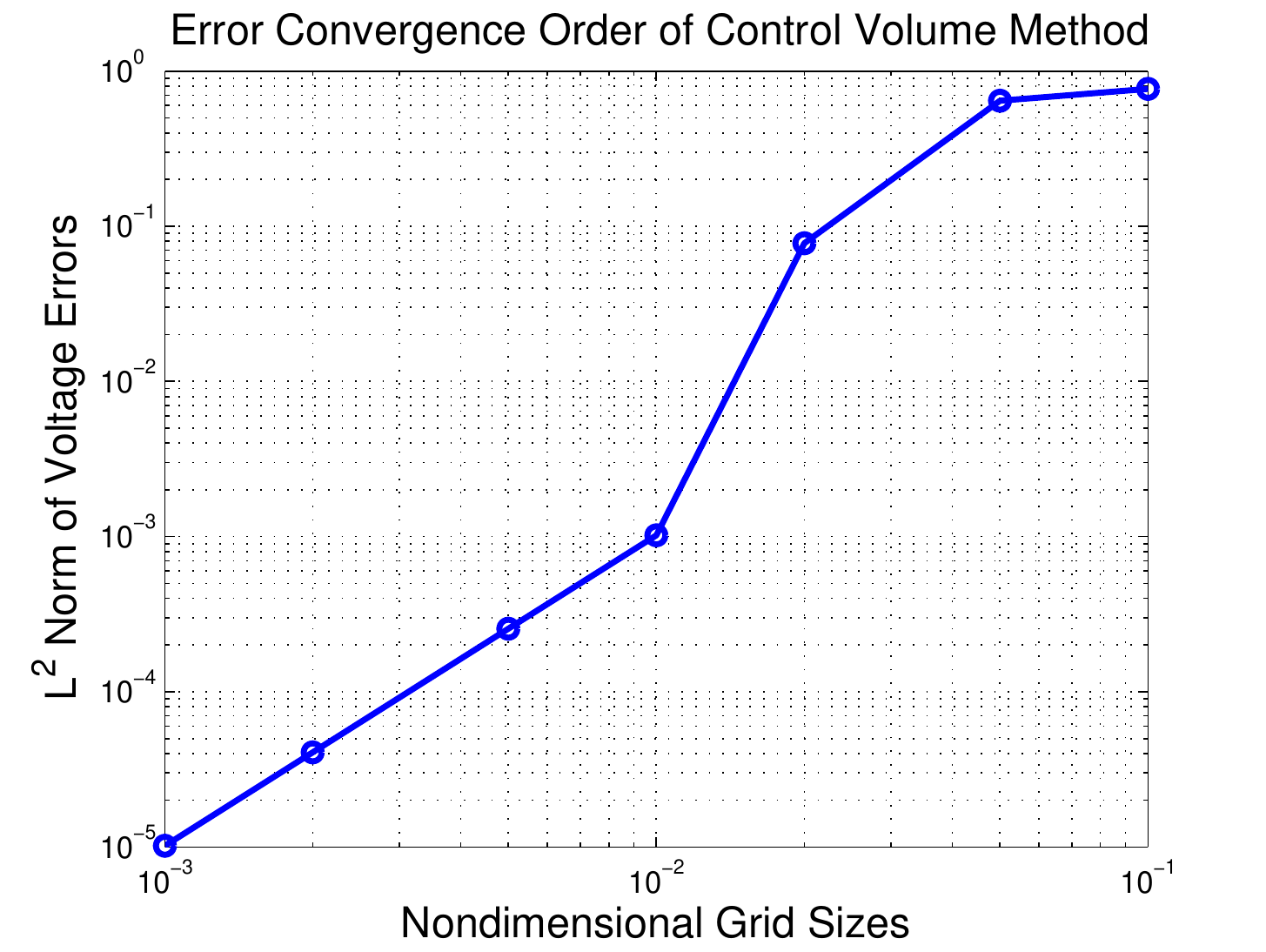} & \includegraphics[width=0.5 \columnwidth]{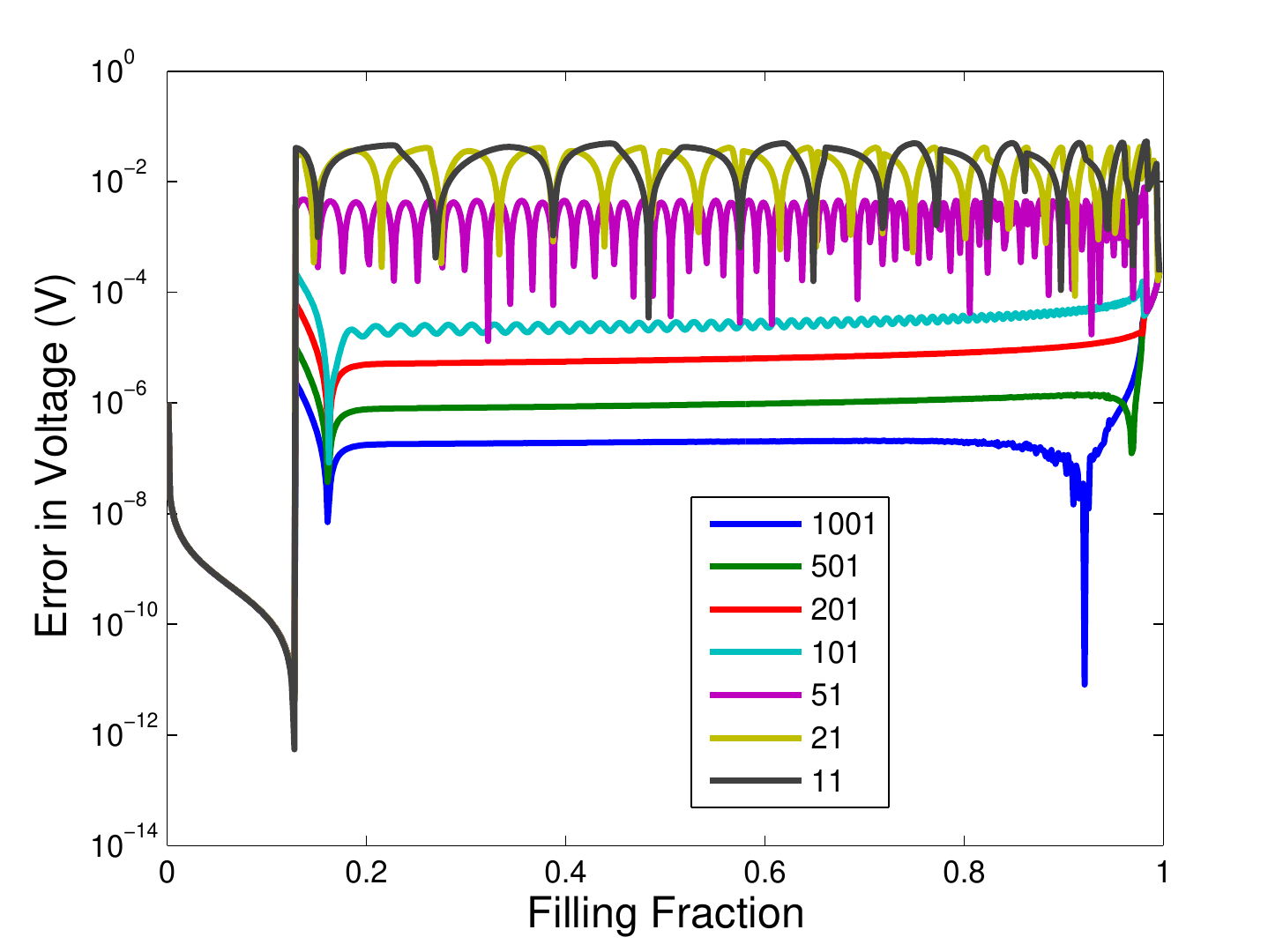}
\end{tabular}
\end{center}
\caption{Error convergence test with the very small current density \textcolor{black}{ $10^{-4}$C,} while no surface wetting is assumed. The error is defined as the $l^2$ norm of the voltage vector difference from the reference solution over the square root of length of this vector. The error converges in second order as suggested by the figure on the left. We also plot the error in voltage during ion intercalation of all these grid point cases (solution from 11 points to 1001 points compare to the reference solution from 3001 grids) in the right figure, where we observe oscillations when the grid is coarse.}
\label{fig:ErrorConvergence}
\end{figure}

The plot of error convergence is shown in the left half of Fig. \ref{fig:ErrorConvergence}, which is consistent with our previous expectation. The absolute error in voltage shown in the right hand side in the same figure signifies that we will have trouble with oscillations after the phase separation if the grid is not fine enough.

\begin{figure}[!h]
\begin{center}
\begin{tabular}{ c c}
  \includegraphics[width=0.5 \columnwidth]{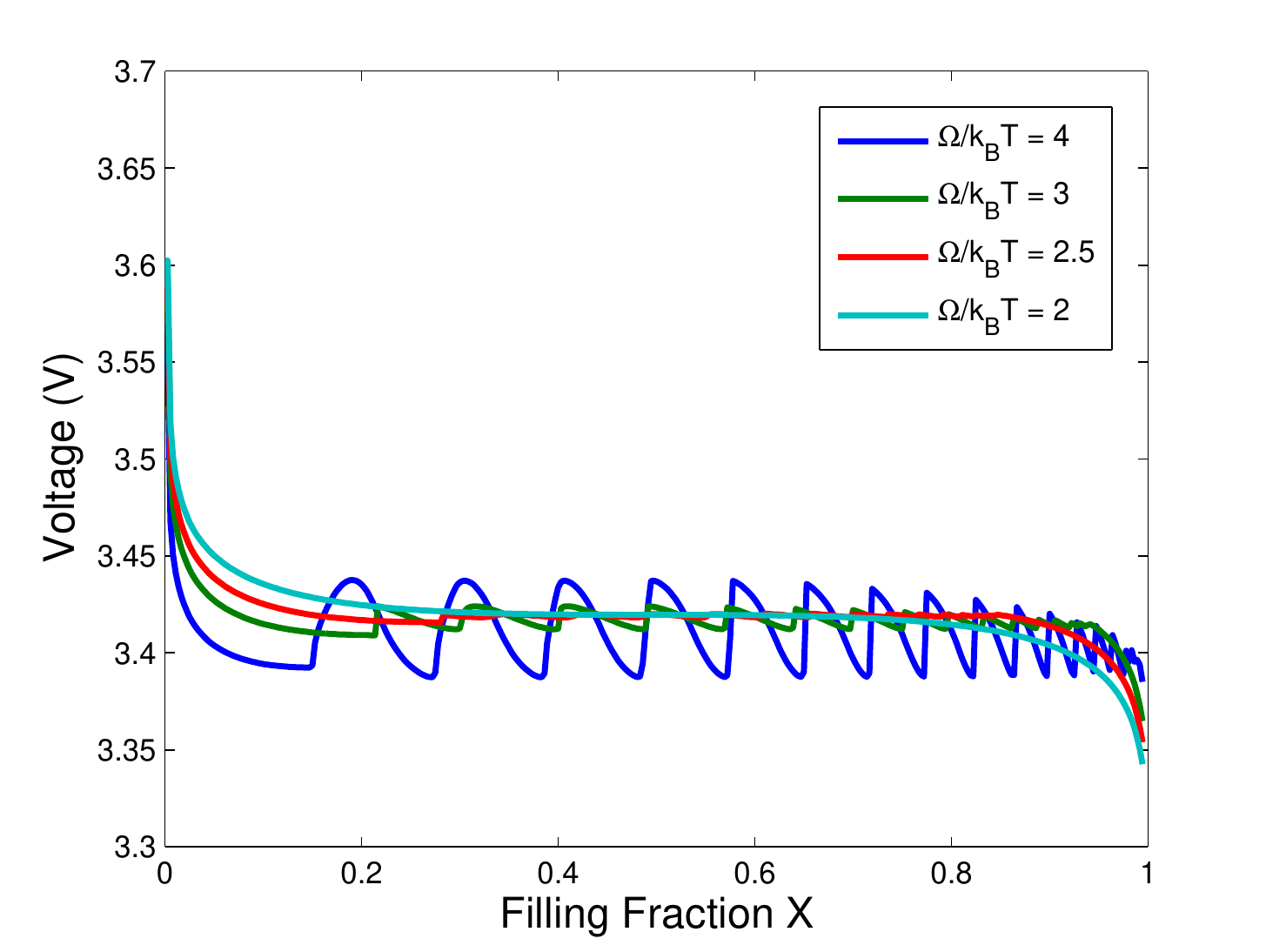} & \includegraphics[width=0.5 \columnwidth]{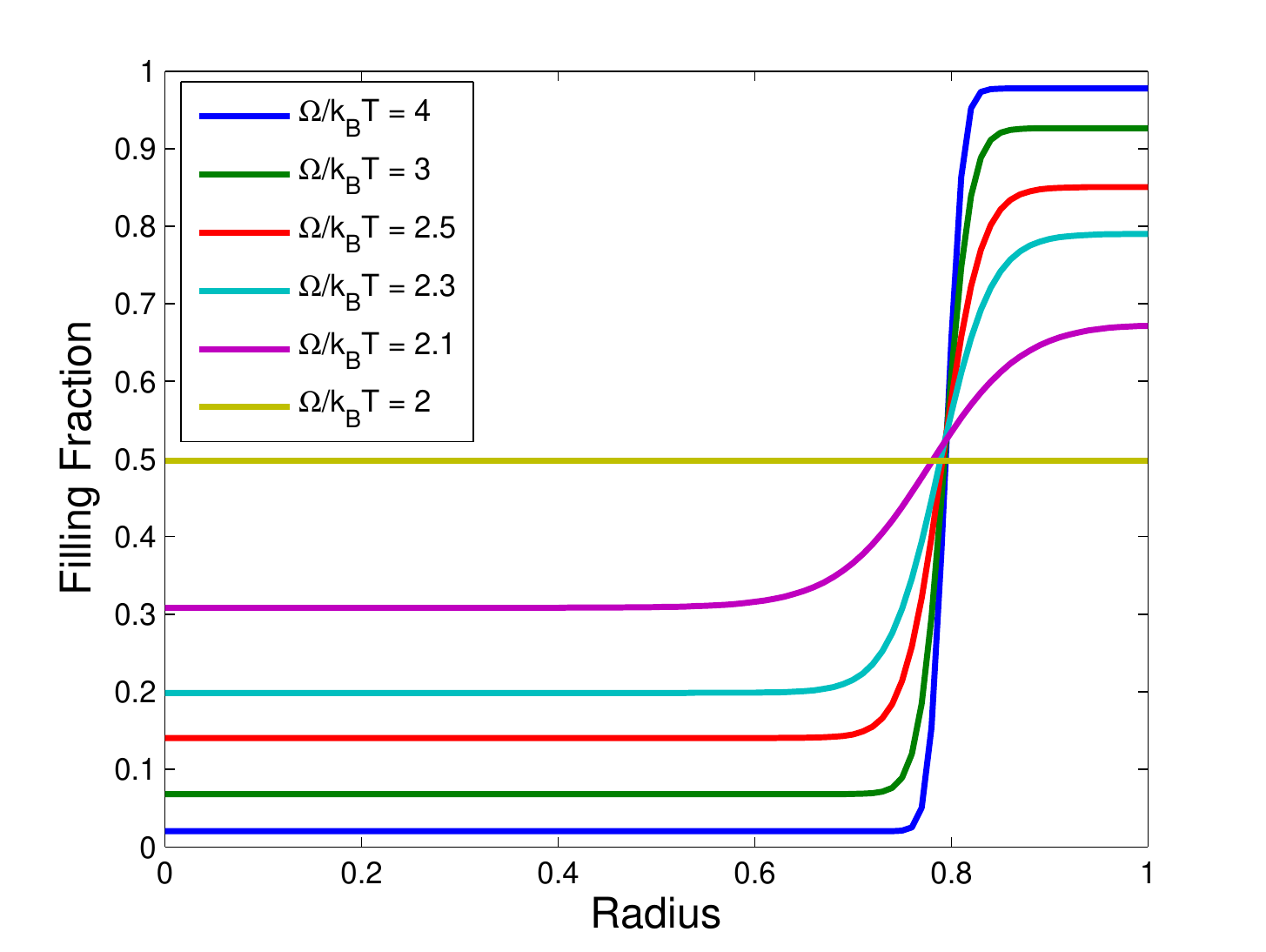} \\   
\end{tabular}
\end{center}
\caption{Voltage prediction plot with different $\tilde \Omega$ using 21 grid points on the left. We see more oscillations in larger $\tilde \Omega$. The right hand side is the concentration distribution with different $\tilde \Omega$ when the \textcolor{black}{filling fraction $X = \frac{1}{2}$}. Higher $\tilde \Omega$ value indicates a thinner phase boundary thickness. The current density is set to be \textcolor{black}{ $10^{-4}$C,} and no surface wetting is assumed in both of these simulations.}
\label{fig:Oscillation}
\end{figure}

As we see from Fig. \ref{fig:Oscillation}, with $21$ grid points, we may get different oscillation sizes in the solutions, which is sensitive to the parameter $\tilde \Omega$. While \textcolor{black}{comparing} to the concentration distribution on the right, a larger parameter $\tilde \Omega$ leads to a smaller interfacial width, we need a fine enough grid which is with the grid size  smaller than the interfacial width to capture the propagating shock without creating oscillations.

Therefore, in the choice of grid point number, we need to be careful about all conditions such as the radius, $\tilde \Omega$ and $\kappa$ in order to get the desired accuracy with good stability, but without paying too much for the computation cost.

\section{Conclusion}
\label{sec:conc}

In summary, we have studied the dynamics of ion intercalation in an isotropic spherical battery intercalation particle using the heterogeneous  CHR model with Butler-Volmer reaction kinetics ~\cite{bazant2013}.  The model predicts either solid solution with radial nonlinear diffusion or  core-shell phase separation, depending on the thermodynamic, geometrical, and electrochemical conditions. The model is able to consistently predict the transient voltage after a current step, regardless of the complexity of the dynamics, far from equilibrium.  Surface wetting plays a major role in nucleating phase separation.  The simplifying assumptions of radial symmetry and negligible coherency strain maybe be applicable to some materials, such as lithium titanate anodes or defective lithium iron phosphate cathodes, while the basic principles illustrated here have broad relevance for intercalation materials with complex thermodynamics and multiple stable phases.

\section*{Acknowledgments}

This material is based upon work supported by the National Science Foundation Graduate Research Fellowship under Grant No. 1122374.  This work was also partially supported by the Samsung-MIT Alliance. \textcolor{black}{We thank Peng Bai for useful discussions and the anonymous reviewers for their helpful suggestions on this paper.}

\renewcommand\refname{Reference}
\bibliographystyle{siam}
\bibliography{elec42}

\begin{thebibliography}{10}

\bibitem{allen2008}
{\sc JL~Allen, TR~Jow, and J~Wolfenstine}, {\em Analysis of the
  \(\mbox{FePO}_4\) to \(\mbox{LiFePO}_4\) phase transition}, Journal of Solid
  State Electrochemistry, 12 (2008), pp.~1031--1033.

\bibitem{bai2014}
{\sc P.~Bai and M.~Z. Bazant}, {\em Charge transfer kinetics at the solid-solid
  interface in porous electrodes},  (2014).
\newblock submitted.

\bibitem{bai2011}
{\sc P.~Bai, D.~A. Cogswell, and Martin~Z. Bazant}, {\em Suppression of phase
  separation in \(\mbox{LiFePO}_4\) nanoparticles during battery discharge},
  Nano Letters, 11 (2011), pp.~4890--4896.

\bibitem{bai2013}
{\sc Peng Bai and Guangyu Tian}, {\em Statistical kinetics of
  phase-transforming nanoparticles in $\mbox{LiFePO}_4$ porous electrodes},
  Electrochimica Acta, 89 (2013), pp.~644--651.

\bibitem{kom}
{\sc Robert~W. Balluffi, Samuel~M. Allen, and W.~Craig Carter}, {\em Kinetics
  of materials}, Wiley, 2005.

\bibitem{Banas2008}
{\sc Lubomir Banas and Robert N{\"u}rnberg}, {\em Adaptive finite element
  methods for cahn--hilliard equations}, Journal of Computational and Applied
  Mathematics, 218 (2008), pp.~2--11.

\bibitem{bard_book}
{\sc A.~J. Bard and L.~R. Faulkner}, {\em Electrochemical Methods}, J. Wiley \&
  Sons, Inc., New York, NY, 2001.

\bibitem{barenblatt_book}
{\sc G.I. Barenblatt}, {\em Similarity, Self-Similarity and Intermediate
  Asymptotics}, Cambridge University Press, 2nd edition~ed., 1996.

\bibitem{10.626}
{\sc M.~Z. Bazant}, {\em 10.626 Electrochemical Energy Systems}, Massachusetts
  Institute of Technology: MIT OpenCourseWare, http://ocw.mit.edu, License:
  Creative Commons BY-NC-SA, 2011.
\newblock Lecture 11.

\bibitem{bazant2013}
\leavevmode\vrule height 2pt depth -1.6pt width 23pt, {\em Theory of chemical
  kinetics and charge transfer based on non-equilibrium thermodynamics},
  Accounts of Chemical Research, 46 (2013), pp.~1144--1160.

\bibitem{burch_thesis}
{\sc Damian Burch}, {\em Intercalation Dynamics in Lithium-Ion Batteries},
  Ph.D. Thesis in Mathematics, Massachusetts Institute of Technology, 2009.

\bibitem{burch2009}
{\sc Damian Burch and Martin~Z. Bazant}, {\em Size-dependent spinodal and
  miscibility gaps for intercalation in nanoparticles}, Nano Letters, 9 (2009),
  pp.~3795--3800.

\bibitem{cahn1959-1}
{\sc John Cahn}, {\em Free energy of a nonuniform system. ii. thermodynamic
  basis}, Journal of Chemical Physics, 30 (1959), pp.~1121--1124.

\bibitem{cahn1959-2}
{\sc John Cahn and John Hilliard}, {\em Free energy of a nonuniform system.
  iii. nucleation in a two-component incompressible fluid}, Journal of Chemical
  Physics, 31 (1959), pp.~688--699.

\bibitem{cahn1977}
{\sc J.~W. Cahn}, {\em Critical point wetting}, J. Chem. Phys., 66 (1977),
  pp.~3667--3672.

\bibitem{cahn1958}
{\sc J.~W. Cahn and J.~W. Hilliard}, {\em Free energy of a non-uniform system:
  I. interfacial energy}, J. Chem Phys., 28 (1958), pp.~258--267.

\bibitem{chen2006}
{\sc Guoying Chen, Xiangyun Song, and Thomas Richardson}, {\em Electron
  microscopy study of the \(\mbox{LiFePO}_4\) to \(\mbox{FePO}_4\) phase
  transition}, Electrochemical and Solid State Letters, 9 (2006),
  pp.~A295--A298.

\bibitem{choo1998}
{\sc SM~Choo and SK~Chung}, {\em Conservative nonlinear difference scheme for
  the cahn-hilliard equation}, Computers \& Mathematics with Applications, 36
  (1998), pp.~31--39.

\bibitem{chueh2013}
{\sc William~C Chueh, Farid~El Gabaly, Josh~D Sugar, Norman~C. Bartelt,
  Anthony~H. McDaniel, Kyle~R Fenton, Kevin~R. Zavadil, Tolek Tyliszczak, Wei
  Lai, and Kevin~F. McCarty}, {\em Intercalation pathway in many-particle
  $\mbox{LiFePO}_4$ electrode revealed by nanoscale state-of-charge mapping},
  Nano Letters, 13 (2013), pp.~866 -- 872.

\bibitem{cogswell2012}
{\sc Daniel~A. Cogswell and Martin~Z. Bazant}, {\em Coherency strain and the
  kinetics of phase separation in $\mbox{LiFePO}_4$ nanoparticles}, ACS Nano, 6
  (2012), pp.~2215--2225.

\bibitem{cogswell2013}
{\sc D.~A. Cogswell and M.~Z. Bazant}, {\em Theory of coherent nucleation in
  phase-separating nanoparticles}, Nano Letters, 13 (2013), pp.~3036--3041.

\bibitem{cueto2008}
{\sc Luis Cueto-Felgueroso and Jaume Peraire}, {\em A time-adaptive finite
  volume method for the cahn-hilliard and kuramoto--sivashinsky equations},
  Journal of Computational Physics, 227 (2008), pp.~9985--10017.

\bibitem{dargaville_thesis}
{\sc Steven Dargaville}, {\em Mathematical Modelling of $\mbox{LiFePO}_4$
  Cathodes}, Ph.D. Thesis in Mathematics, Queensland University of Technology,
  2013.

\bibitem{dargaville2010}
{\sc S.~Dargaville and T.W. Farrell}, {\em Predicting active material
  utilization in \(\mbox{LiFePO}_4\) electrodes using a multiscale mathematical
  model}, Journal of the Electrochemical Society, 157 (2010), pp.~A830--A840.

\bibitem{dargaville2013numerical}
{\sc Steven Dargaville and Troy~W Farrell}, {\em A least squares based finite
  volume method for the cahn-hilliard and cahn-hilliard-reaction equations},
  Journal of Computational and Applied Mathematics, in press (2013).

\bibitem{dargaville2013CHR}
{\sc S.~Dargaville and T.~W. Farrell}, {\em The persistence of phase-separation
  in lifepo4 with two-dimensional li+ transport: The cahn-hilliard-reaction
  equation and the role of defects}, Electrochimica Acta, 94 (2013),
  pp.~143--158.

\bibitem{de2005}
{\sc EVL De~Mello and Otton Teixeira~da Silveira~Filho}, {\em Numerical study
  of the cahn--hilliard equation in one, two and three dimensions}, Physica A:
  Statistical Mechanics and its Applications, 347 (2005), pp.~429--443.

\bibitem{dehghan2009}
{\sc Mehdi Dehghan and Davoud Mirzaei}, {\em A numerical method based on the
  boundary integral equation and dual reciprocity methods for one-dimensional
  cahn--hilliard equation}, Engineering analysis with boundary elements, 33
  (2009), pp.~522--528.

\bibitem{delacourt2005}
{\sc Charles Delacourt, Philippe Poizot, Jean-Marie Tarascon, and Christian
  Masquelier}, {\em The existence of a temperature-driven solid solution in
  \(\mbox{Li$_x$FePO}_4\) for 0 $\leq$ x $\leq$ 1}, Nature materials, 4 (2005),
  pp.~254--260.

\bibitem{delmas2008}
{\sc C.~Delmas, M.~Maccario, L.~Croguennec, F.~Le Cras, and F.~Weill}, {\em
  Lithium deintercalation of $\mbox{LiFePO}_4$ nanoparticles via a
  domino-cascade model}, Nature Materials, 7 (2008), pp.~665--671.

\bibitem{vanderven2009}
{\sc A.~Van der Ven, K.~Garikipati, S.~Kim, and M.~Wagemaker}, {\em The role of
  coherency strains on phase stability in li$_x$fepo$_4$: Needle crystallites
  minimize coherency strain and overpotential}, J. Electrochem. Soc., 156
  (2009), pp.~A949--A957.

\bibitem{doyle1993}
{\sc Marc Doyle, Thomas~F. Fuller, and John Newman}, {\em Modeling of
  galvanostatic charge and discharge of the lithium/polymer/insertion cell},
  Journal of the Electrochemical Society, 140 (1993), pp.~1526--1533.

\bibitem{dreyer2011}
{\sc D.~Dreyer, C.~Guhlke, and R.~Huth}, {\em The behavior of a many-particle
  electrode in a lithium-ion battery}, Physica D, 240 (2011), pp.~1008--1019.

\bibitem{dreyer2010}
{\sc Wolfgang Dreyer, Janko Jamnik, Clemens Guhlke, Robert Huth, Joze Moskon,
  and Miran Gaberscek}, {\em The thermodynamic origin of hysteresis in
  insertion batteries}, Nat. Mater., 9 (2010), pp.~448--453.

\bibitem{fehribach2009}
{\sc J.~D. Fehribach and R.~O'Hayre}, {\em Triple phase boundaries in
  solid-oxide cathodes}, SIAM J. Appl. Math., 70 (2009), pp.~510--530.

\bibitem{ferguson2012}
{\sc T.~R. Ferguson and M.~Z. Bazant}, {\em Non-equilibrium thermodynamics of
  porous electrodes}, J. Electrochem. Soc., 159 (2012), pp.~A1967--A1985.

\bibitem{ferguson2014}
\leavevmode\vrule height 2pt depth -1.6pt width 23pt, {\em Phase transformation
  dynamics in porous battery electrodes},  (2014).
\newblock submitted, arXiv:1401.7072 [physics.chem-ph].

\bibitem{greer2006}
{\sc John~B Greer, Andrea~L Bertozzi, and Guillermo Sapiro}, {\em Fourth order
  partial differential equations on general geometries}, Journal of
  Computational Physics, 216 (2006), pp.~216--246.

\bibitem{han2004}
{\sc B.C. Han, A.~Van der Ven, D.~Morgan, and G.~Ceder}, {\em Electrochemical
  modeling of intercalation processes with phase field models}, Electrochimica
  Acta, 49 (2004), pp.~4691--4699.

\bibitem{he2009}
{\sc Li-ping He and Yunxian Liu}, {\em A class of stable spectral methods for
  the cahn--hilliard equation}, Journal of Computational Physics, 228 (2009),
  pp.~5101--5110.

\bibitem{hildebrand2003}
{\sc M.~Hildebrand, M.~Ipsen, A.~S. Mikhailov, and G~Ertl}, {\em Localized
  nonequilibrium nanostructures in surface chemical reactions}, New J. Phys., 5
  (2003), pp.~61.1--61.28.

\bibitem{hildebrand1999}
{\sc M.~Hildebrand, M.~Kuperman, H.~Wio, A.~S. Mikhailov, and G.~Ertl}, {\em
  Self-organized chemical nanoscale microreactors}, Phys. Rev. Lett., 83
  (1999), pp.~1475--1478.

\bibitem{horstmann2013}
{\sc B.~Horstmann, B.~Gallant, R.~Mitchell, W.~G. Bessler, Y.~Shao-Horn, and
  M.~Z. Bazant}, {\em Rate-dependent morphology of li$_2$o$_2$ growth in
  li--o$_2$ batteries}, J. Phys. Chem. Lett., 4 (2013), pp.~4217--4222.

\bibitem{kang2009}
{\sc Byoungwoo Kang and Gerbrand Ceder}, {\em Battery materials for ultrafast
  charging and discharging}, Nature, 458 (2009), pp.~190--193.

\bibitem{kao2010}
{\sc Yu-Hua Kao, Ming Tang, Nonglak Meethong, Jianming Bai, W.~Craig Carter,
  and Yet-Ming Chiang}, {\em Overpotential-dependent phase transformation
  pathways in lithium iron phosphate battery electrodes}, Chem. Mater., 22
  (2010), pp.~5845--5855.

\bibitem{kasavajjula2008}
{\sc Uday~S. Kasavajjula, Chunsheng Wang, and Pedro~E. Arce}, {\em Discharge
  model for lifepo4 accounting for the solid solution range}, J. Electrochem.
  Soc., 155 (2008), pp.~A866--A874.

\bibitem{kim2004}
{\sc Junseok Kim, Kyungkeun Kang, and John Lowengrub}, {\em Conservative
  multigrid methods for cahn--hilliard fluids}, Journal of Computational
  Physics, 193 (2004), pp.~511--543.

\bibitem{laffont2006}
{\sc L.~Laffont, C.~Delacourt, P.~Gibot, M.~Yue Wu, P.~Kooyman, C.~Masquelier,
  and J.~Marie Tarascon}, {\em Study of the $\mbox{LiFePO}_4$/$\mbox{FePO}_4$
  two-phase system by high-resolution electron energy loss spectroscopy}, Chem.
  Mater., 18 (2006), pp.~5520--5529.

\bibitem{lai2011b}
{\sc Wei Lai}, {\em Electrochemical modeling of single particle intercalation
  battery materials with different thermodynamics}, Journal of Power Sources,
  196 (2011), pp.~6534--6553.

\bibitem{lai2010}
{\sc W.~Lai and F.~Ciucci}, {\em Thermodynamics and kinetics of phase
  transformation in intercalation battery electrodes - phenomenological
  modeling}, Electrochim. Acta, 56 (2010), pp.~531--542.

\bibitem{lai2011a}
{\sc Wei Lai and Francesco Ciucci}, {\em Mathematical modeling of porous
  battery electrodes - revisit of newman's model}, Electrochimica Acta, 56
  (2011), pp.~4369--4377.

\bibitem{malik2010}
{\sc Rahul Malik, Damian Burch, Martin Bazant, and Gerbrand Ceder}, {\em
  Particle size dependence of the ionic diffusivity}, Nano Letters, 10 (2010),
  pp.~4123--4127.

\bibitem{meethong2007}
{\sc Nonglak Meethong, Hsiao-Ying~Shadow Huang, W.~Craig Carter, and Yet-Ming
  Chiang}, {\em Size-dependent lithium miscibility gap in nanoscale
  li$_{1-x}$fepo$_4$}, Electrochem. Solid-State Lett., 10 (2007),
  pp.~A134--A138.

\bibitem{meethong2007a}
{\sc N.~Meethong, H.~Y.~S. Huang, S.~A. Speakman, W.~C. Carter, and Y.~M.
  Chiang}, {\em Strain accommodation during phase transformations in
  olivine-based cathodes as a materials selection criterion for high-power
  rechargeable batteries}, Adv. Funct. Mater., 17 (2007), pp.~1115--1123.

\bibitem{meethong2008}
{\sc Nonglak Meethong, Yu-Hua Kao, Ming Tang, Hsiao-Ying Huang, W.~Craig
  Carter, and Yet-Ming Chiang}, {\em Electrochemically induced phase
  transformation in nanoscale olivines li$_(1-x)$mpo4 (m = fe, mn)}, Chem.
  Mater., 20 (2008), pp.~6189--6198.

\bibitem{morgan2004}
{\sc D.~Morgan, A.~Van der Ven, and G.~Ceder}, {\em Li conductivity in
  $\mbox{Li}_x\mbox{MPO}_4$ (m=mn,fe,co,ni) olivine materials}, Electrochemical
  and Solid State Letters, 7 (2004), pp.~A30--A32.

\bibitem{nauman2001}
{\sc E.~Bruce Nauman and D.~Qiwei Heb}, {\em Nonlinear diffusion and phase
  separation}, Chemical Engineering Science, 56 (2001), pp.~1999--2018.

\bibitem{newman_book}
{\sc John Newman and Karen~E. Thomas-Alyea}, {\em Electrochemical Systems},
  Prentice-Hall, Inc., Englewood Cliffs, NJ, third~ed., 2004.

\bibitem{ozhuku1995}
{\sc Tsutamu Ohzuku, Atsushi Ueda, and Norihiro Yamamota}, {\em Zero-strain
  insertion material of li[lil/3tis/3]o4 for rechargeable lithium cells}, J.
  Electrochem. Soc., 142 (1995), pp.~1431--1435.

\bibitem{orvananos2014}
{\sc Bernardo Orvananos, Todd~R. Ferguson, Hui-Chia Yu, Martin~Z. Bazant, and
  Katsuyo Thornton}, {\em Particle-level modeling of the charge-discharge
  behavior of nanoparticulate phase-separating li-ion battery electrodes}, J.
  Electrochem. Soc., in press (2014).
\newblock arXiv:1309.6495 [cond-mat.mtrl-sci].

\bibitem{oyama2012}
{\sc Gosuke Oyama, Yuki Yamada, Ryuichi Natsui, Shinichi Nishimura, and Atsuo
  Yamada}, {\em Kinetics of nucleation and growth in two-phase electrochemical
  reaction of $\mbox{LiFePO}_4$}, J. Phys. Chem. C, 116 (2012), pp.~7306--7311.

\bibitem{padhi1997}
{\sc A.K. Padhi, K.S. Nanjundaswamy, and J.B. Goodenough}, {\em
  Phospho-olivines as positive-electrode materials for rechargeable lithium
  batteries}, Journal of the Electrochemical Society, 144 (1997),
  pp.~1188--1194.

\bibitem{ritchie2006}
{\sc Andrew Ritchie and Wilmont Howard}, {\em Recent developments and likely
  advances in lithium-ion batteries}, Journal of Power Sources, 162 (2006),
  pp.~809--812.

\bibitem{shin2011}
{\sc Jaemin Shin, Darae Jeong, and Junseok Kim}, {\em A conservative numerical
  method for the cahn--hilliard equation in complex domains}, Journal of
  Computational Physics, 230 (2011), pp.~7441--7455.

\bibitem{singh2008}
{\sc Gogi Singh, Damian Burch, and Martin~Z. Bazant}, {\em Intercalation
  dynamics in rechargeable battery materials: General theory and
  phase-transformation waves in \(\mbox{LiFePO}_4\)}, Electrochimica Acta, 53
  (2008), pp.~7599--7613.
\newblock arXiv:0707.1858v1 [cond-mat.mtrl-sci] (2007).

\bibitem{srinivasan2004}
{\sc Venkat Srinivasan and John Newman}, {\em Discharge model for the lithium
  iron-phosphate electrode}, Journal of the Electrochemical Society, 151
  (2004), pp.~A1517--A1529.

\bibitem{tang2011}
{\sc Ming Tang, James~F. Belak, and Milo~R. Dorr}, {\em Anisotropic phase
  boundary morphology in nanoscale olivine electrode particles}, The Journal of
  Physical Chemistry C, 115 (2011), pp.~4922--4926.

\bibitem{tang2010}
{\sc Ming Tang, W.~Craig Carter, and Yet-Ming Chiang}, {\em Electrochemically
  driven phase transitions in insertion electrodes for lithium-ion batteries:
  Examples in lithium metal phosphate olivines}, Annual Review of Materials
  Research, 40 (2010), pp.~501--529.

\bibitem{tang2009}
{\sc M.~Tang, H.-Y. Huang, N.~Meethong, Y.-H. Kao, W.~C. Carter, and Y.-M.
  Chiang}, {\em Model for the particle size, overpotential, and strain
  dependence of phase transition pathways in storage electrodes: Application to
  nanoscale olivines}, Chem. Mater., 21 (2009), pp.~1557--1571.

\bibitem{tarascon2001}
{\sc J.M. Tarascon and M.~Armand}, {\em Issues and challenges facing
  rechargeable lithium batteries}, Nature, 414 (2001), pp.~359--367.

\bibitem{wagemaker2011}
{\sc Marnix Wagemaker, Deepak~P. Singh, Wouter~J.H. Borghols, Ugo Lafont, Lucas
  Haverkate, Vanessa~K. Peterson, and Fokko~M. Mulder}, {\em Dynamic solubility
  limits in nanosized olivine $\mbox{LiFePO}_4$}, J. Am. Chem. Soc., 133
  (2011), pp.~10222--10228.

\bibitem{cwang2007}
{\sc Chunsheng Wang, Uday~S. Kasavajjula, and Pedro~E. Arce}, {\em A discharge
  model for phase transformation electrodes:? formulation, experimental
  validation, and analysis}, J. Phys. Chem. C, 111 (2007), pp.~16656--16663.

\bibitem{wise2007}
{\sc Steven Wise, Junseok Kim, and John Lowengrub}, {\em Solving the
  regularized, strongly anisotropic cahn--hilliard equation by an adaptive
  nonlinear multigrid method}, Journal of Computational Physics, 226 (2007),
  pp.~414--446.

\bibitem{wodo2011}
{\sc Olga Wodo and Baskar Ganapathysubramanian}, {\em Computationally efficient
  solution to the cahn--hilliard equation: Adaptive implicit time schemes, mesh
  sensitivity analysis and the 3d isoperimetric problem}, Journal of
  Computational Physics, 230 (2011), pp.~6037--6060.

\bibitem{xia2007}
{\sc Yinhua Xia, Yan Xu, and Chi-Wang Shu}, {\em Local discontinuous galerkin
  methods for the cahn--hilliard type equations}, Journal of Computational
  Physics, 227 (2007), pp.~472--491.

\bibitem{yamada2005}
{\sc Atsuo Yamada, Hiroshi Koizumi, Noriyuki Sonoyama, and Ryoji Kanno}, {\em
  Phase change in \(\mbox{Li$_x$FePO}_4\)}, Electrochemical and Solid-State
  Letters, 8 (2005), pp.~A409--A413.

\bibitem{yang2009}
{\sc Zhenguo Yang, Daiwon Choi, Sebastien Kerisit, Kevin~M. Rosso, DonghaiWang,
  Jason Zhang, Gordon Graff, and Jun Liu}, {\em Nanostructures and lithium
  electrochemical reactivity of lithium titanites and titanium oxides: A
  review}, J. Power Sources, 192 (2009), pp.~588--598.

\bibitem{zackrisson2010}
{\sc Mats Zackrisson, Lars Avell{\'a}n, and Jessica Orlenius}, {\em Life cycle
  assessment of lithium-ion batteries for plug-in hybrid electric
  vehicles--critical issues}, Journal of Cleaner Production, 18 (2010),
  pp.~1519--1529.

\bibitem{zeng2013numerical}
{\sc Yi~Zeng, Paul Albertus, Reinhardt Klein, Nalin Chaturvedi, Aleksandar
  Kojic, Martin~Z Bazant, and Jake Christensen}, {\em Efficient conservative
  numerical schemes for 1d nonlinear spherical diffusion equations with
  applications in battery modeling}, Journal of The Electrochemical Society,
  160 (2013), pp.~A1565--A1571.

\bibitem{zeng2013MRS}
{\sc Y.~Zeng and M.~Z. Bazant}, {\em Cahn-hilliard reaction model for isotropic
  li-ion battery nanoparticles}, MRS Proceedings, 1542 (2013).

\bibitem{zhang2010}
{\sc Shuo Zhang and Ming Wang}, {\em A nonconforming finite element method for
  the cahn--hilliard equation}, Journal of Computational Physics, 229 (2010),
  pp.~7361--7372.

\end{thebibliography}

\end{document}